\documentclass[sigconf]{acmart}
\renewcommand\footnotetextcopyrightpermission[1]{} % removes footnote with conference information in first column
\pdfoutput=1
\setcopyright{none}
\usepackage{color} 
\usepackage{booktabs} % For formal tables
\newenvironment{itemize*}{\begin{itemize}}{\end{itemize}}
\usepackage{enumitem}
\usepackage[utf8x]{inputenc}
\usepackage{subfigure}
\usepackage{graphicx}
\usepackage{balance}
\usepackage{comment}
\usepackage{amsmath}
\usepackage{xspace}
\usepackage{caption}
\usepackage{siunitx}
\usepackage{booktabs}
\usepackage{enumitem}
\setlist[itemize]{leftmargin=*,noitemsep}

\newcommand{\system}{\textsc{LoRea}\xspace}

\newcommand{\fakeparagraph}[1]{\vspace{.5mm}\noindent\textbf{#1.}}
\newcommand{\fakepar}[1]{\fakeparagraph{#1}}
\newcommand{\fakeparagraphnodot}[1]{\vspace{.5mm}\noindent\textbf{#1}}
\newcommand{\fakeparnodot}[1]{\fakeparagraphnodot{#1}}
\newcommand{\capt}[1]{\mdseries{\emph{#1}}}


\begin{document}

%\title{\system: A  Backscatter Architecture that Achieves a Long Communication Range}
\title{LoRea: A  Backscatter Architecture that Achieves a Long Communication Range}

\author{Ambuj Varshney}
\affiliation{%
  \institution{Uppsala University}
  \city{ Sweden} 
}
\email{ambuj.varshney@it.uu.se}

\author{Oliver Harms}
\affiliation{%
  \institution{Uppsala University}
  \city{ Sweden} 
}
\email{mail@oliverharms.eu}

\author{Carlos Pérez-Penichet}
\affiliation{%
  \institution{Uppsala University}
  \city{ Sweden} 
}
\email{carlos.penichet@it.uu.se}

\author{Christian Rohner}
\affiliation{%
  \institution{Uppsala University}
  \city{ Sweden} 
}
\email{christian.rohner@it.uu.se}

\author{Frederik Hermans}
\affiliation{%
  \institution{Uppsala University}
  \city{ Sweden} 
}
\email{frederik@it.uu.se}

\author{Thiemo Voigt}
\affiliation{%
  \institution{Uppsala University and RISE SICS}
  \city{Sweden} 
}
\email{thiemo@sics.se}

\renewcommand{\shortauthors}{Varshney et al.}

\begin{abstract}
There is the long-standing assumption 
that radio communication
in the  range of hundreds of meters needs
to consume mWs of power at the transmitting device. 
In this paper, we demonstrate that this is 
not necessarily the case for some devices
equipped with backscatter radios.
We present \system an architecture consisting of
a  tag, a reader and multiple carrier generators
that overcomes the power, cost and range limitations
of existing systems such as  Computational Radio Frequency Identification~(CRFID). \system achieves this by: \emph{First}, generating
narrow-band backscatter transmissions
that improve receiver sensitivity.
\emph{Second}, mitigating
self-interference  without the complex designs
employed on RFID readers by  
keeping carrier signal and backscattered
signal apart in frequency.  \emph{Finally}, decoupling 
carrier generation from the reader and using devices  such as 
WiFi routers and sensor nodes as a source of the carrier signal.
An off-the-shelf  implementation of \system costs 70 USD, a drastic reduction in price considering commercial RFID readers
cost 2000 USD. \system's 
range scales with the carrier strength, and proximity to
the carrier source and achieves a maximum 
range of \SI{3.4}{\kilo\meter} 
when the tag is located at
\SI{1}{\meter} distance from a \SI{28}{\decibel}m carrier
source while consuming \SI{70}{\micro\watt} at the tag. When the tag is equidistant from the carrier source 
and the receiver, we can communicate upto \SI{75}{\meter}, a significant improvement over existing RFID
readers.

\end{abstract}

\maketitle

\section{Introduction}

Backscatter communication enables wireless transmissions at a power consumption
orders of magnitude lower than traditional radios. 
A backscatter transmitter modulates ambient 
wireless signals by selectively reflecting 
or absorbing them, which consumes 
less than \SI{1}{\micro\watt} of power~\cite{ambientbackscatter}. This
makes backscatter communications well-suited for
applications 
where replacing batteries is challenging~\cite{wispcam,yeager2009neuralwisp}
or where extending battery life is important~\cite{braidio}. In the past few years,
significant progress has been made to advance backscatter communication.
Recent works demonstrate the ability to synthesise transmissions compatible
with WiFi (802.11b)~\cite{kellogg2016passive}, BLE~\cite{blebackscatter} and ZigBee~\cite{interscatter,passivezigbee} at \SI{}{\micro\watt}s of power using backscatter transmissions.
Other works leverage ambient wireless signals like television~\cite{ambientbackscatter,turbochargebackscatter} or WiFi~\cite{HitchHike,wifibackscatter,zhang_enabling_2016}
for communication.  On the other hand, 
 the design
of traditional backscatter readers and tags, 
e.g., CRFID systems,  has 
not seen major improvements despite their continuing significance~\cite{hu2015laissez,chain,wispcam,wisent,zhang2014enabling} and the 
widespread deployment of passive RFID systems. 

Existing CRFIDs, 
like
WISP~\cite{smith2006wirelessly} and
Moo~\cite{zhang2011moo}, 
augment traditional RFID tags with sensing and
computational capabilities~\cite{buettner2011dewdrop}. 
These tags  operate on harvested energy and,
over the years, have been used to prototype many 
applications such as
localisation~\cite{zhao2013battery}, wireless microphones~\cite{talla2013hybrid} or
infrastructure monitoring~\cite{deivasigamani2013review}.  Many of these applications 
require a large communication range, e.g., battery-free cameras~\cite{wispcam}, 
but are restricted to operate at very short range~(few meters) due to
the limited range achievable 
with existing RFID  readers.  Further, these applications are also constrained
by the high cost~( $\geq \$2000$) and power consumption of the readers.

\begin{figure}[!t]
    \centering
    \includegraphics[width=\linewidth]{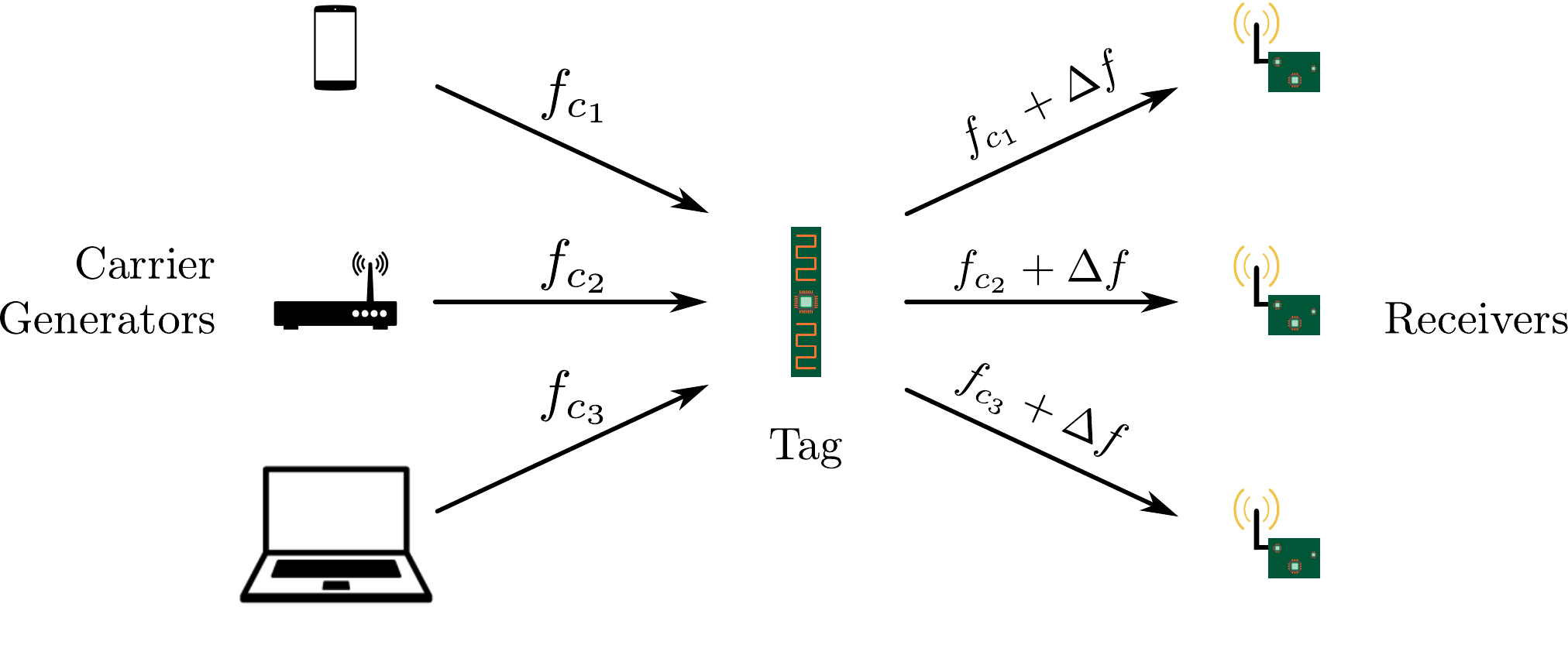}
	\caption{\capt{Overview of our architecture.} One or more devices (sensor nodes, WiFi access points, etc.)
    provide the carrier signal that the backscatter tag reflects to transmit. The backscattered signal is received by one or more receivers.}
    \label{fig:architecture}
\end{figure}

\begin{table*}
\centering
\scriptsize
\begin{tabular}{lrrrrrrrr}
\toprule
\textbf{System name} & \textbf{\system-868} & \textbf{\system-2.4} & Passive WiFi~\cite{kellogg2016passive}  &
HitchHike~\cite{HitchHike}  & InterScatter~\cite{interscatter} &
BLE~\cite{blebackscatter} & BackFi~\cite{BackFi} & RFID~\cite{r420} \\ 
\midrule
\textbf{Carrier strength (dBm)}& 28 & 26  & 30  & 30 & 20 & 15 & 30 & 31.5\\ 
\textbf{Reported range (m)}& 3400  & 225/175 & 33 & 54 & 10 & 9.5 & 5  & $>$10\\ 
\textbf{Bitrate (kbps)} & 2.9 & 2.9/197 & 1000/11000 & 222 & 1000/11000 & 1000 & 1000 & 640\\ 
\textbf{Tag power consumption} & \SI{70}{\micro\watt} &  \SI{650}{\micro\watt} & \SI{14.5}{\micro\watt} (1 Mbps) & \SI{33}{\micro\watt} & \SI{28}{\micro\watt} & N.A & N.A &  30~\SI{}{\micro\watt}\\
							&  & & \SI{59.2}{\micro\watt} (11 Mbps) &  \\
\bottomrule
\end{tabular}
\caption{Comparison of \system with backscatter
    systems which consume \SI{}{\micro\watt}s of power for transmissions. \system's tag was located at a distance of
\SI{1}{m} from the carrier generator, similar to all the other systems. Reported ranges are line-of-sight.}

\label{table:comp}
\end{table*}

To understand the reason for the poor performance of existing CRFID systems,
we see how these systems operate: CRFID tags require an external device~(the reader)
that generates a carrier signal,
provides power, queries  and receives the backscatter reflections from the tags. 
In most CRFID readers, a single device performs all of these operations. The readers receive
backscatter transmissions at the frequency
of the carrier signal~\cite{hu2015laissez,zhang2014ekhonet,
BuZZ, r420,buettner_software_2011}. As energy delivery is combined  
with communication, the readers generate a strong
carrier signal~($\sim$ \SI{30}{dBm}/ \SI{1}{\watt}), which significantly increases their power consumption
making applications such as mobile backscatter readers very challenging
to achieve~(see Section~\ref{mobilereader}). The backscatter reflections,
 are inherently weak, hence separating them from
the strong carrier requires complex techniques 
which increases both cost and complexity~\cite{braidio}.
The readers also suffer from 
poor sensitivity~(-84 dBm~\cite{r420}) due to
leakage of the carrier signal into the receive path~\cite{ma16loc}.

An inexpensive backscatter platform that achieves high communication
range could significantly help  applications conceived using CRFIDs. Further, 
such a platform  could enable new battery-free applications
that are extremely challenging right now. For example, 
sensors embedded within the infrastructure~(see Section~\ref{concrete}).
We present an architecture that attempts to enable  such capability. 
% recently there has been significant 
%interest to develop sensors that can communicate over several hundred
%meters of communication range with several long range wireless standards
%becoming popular. To provide large communication range,
%these applications  use traditional radios which consumes peak power of 
%tens of \SI{}{\milli\watt s}. Such high power consumption required for communication
%forces these sensors  to be reliant batteries for their operations, 
%which increases maintenance cost due to the need to be  replaced often.
%Battery-free operations could significantly help such applications, 
%but existing backscatter tags like
%WISP only demonstrate a range of few meters, while state-of-the-art backscatter systems
%demonstrate range of only tens of meters~\cite{HitchHike, interscatter,kellogg2016passive,blebackscatter}.

\fakepar{Contributions} 
We redesign CRFID-based systems and 
introduce a new architecture shown in Figure~\ref{fig:architecture}. We achieve a
significant improvement across key metrics
like range, price and power consumption in comparison 
to the state of the art~\cite{interscatter,HitchHike,kellogg2016passive,blebackscatter,BackFi}. Our architecture is based
on the following design elements:

\begin{enumerate}[topsep=0pt,itemsep=-1ex,partopsep=1ex,parsep=1ex]

\item  The tradeoff between bitrate and receiver sensitivity is well known. Recent state-of-the-art-
and ultra-low-power backscatter systems
operate at high bitrates (thousands of kilobit/s) due to the use of commodity protocols~\cite{HitchHike,interscatter,kellogg2016passive,blebackscatter,BackFi} which limits
their range and  applicability. We deliberately operate at low bitrates (2.9 kbit/s) which
allows us to use highly sensitive narrow-band receivers. 
Such a design is not detrimental to most sensing applications as they
only send small amounts of information~\cite{oppermann2014decade}. 

\item We  keep the carrier and backscattered signals at different frequencies. 
This improves the SNR of the backscattered signal~(see Section~\ref{sec:background}) by reducing the 
interference from the carrier signal. As opposed to 
traditional readers that use complex solutions to reduce self-interference, our architecture  
leverages the ability of commodity transceivers to reject emissions on adjacent channels.

\item Finally,  we 
use a bistatic configuration where 
the carrier generator and the receiver 
are spatially separated. This has three advantages:
First, spatial separation decreases
self-interference which improves the range owing to path-loss of the carrier
signal. Second,  when operating in the \SI{2.4}{\giga\hertz} band,  we can leverage
commodity devices to provide the carrier signal. Third, decoupling helps to separate the energy-intensive 
carrier generation from the reader.
\end{enumerate}

In our architecture, the communication range scales with the
strength of the carrier signal and the proximity of the tag to the carrier source. 
%which is also a constraint shared by state-of-the-art 
%backscatter systems~\cite{interscatter,kellogg2016passive}.
This property is inherent in state-of-the-art 
backscatter systems~\cite{interscatter,kellogg2016passive}.
When operating in close proximity~(\SI{1}{\meter}), 
and with the strength of the carrier signal close to the maximum permissible power, we 
achieve a range of more than \SI{3.4}{\kilo\meter}  
in the \SI{868}{\mega\hertz} band, and \SI{225}{\meter}
in the \SI{2.4}{\giga\hertz} band with a carrier strength of 
\SI{28}{\decibel}m and \SI{26}{\decibel}m respectively.
This range is an order of magnitude longer than what
state-of-the-art systems achieve~\cite{HitchHike,interscatter,kellogg2016passive} when operating
in similar settings~(see Table~\ref{table:comp}). 
%In a more realistic scenario where
When the tag is located equidistant from both the carrier source and the receiver, a scenario
that encounters path loss similar to monostatic RFID readers, we
achieve a range of \SI{75}{\meter}, a %This  still represents a 
significant improvement
in range over traditional CRFID readers.

Design elements (2) and (3) have also been used in recent backscatter systems~\cite{HitchHike, kellogg2016passive, interscatter}. Combining the three design elements enables us to significantly reduce self-interference without using the complex designs employed by current CRFID readers. This helps us to reduce the price of the reader to 70~USD, a drastic reduction when compared with the approx. 2000 USD that commercial  RFID readers cost~(see Section~\ref{costanalysis}).

Finally, design element (3) enables us to use an infrastructure of wireless
devices as the source of the unmodulated carrier signal. This reduces the power consumption 
of the reader, as the carrier generation is  
the most energy intensive operation 
in backscatter readers. While Interscatter~\cite{interscatter}
demonstrated that BLE radios can be used to generate unmodulated carrier signals, we go a step beyond and demonstrate that 802.15.4 and WiFi
radios can also generate carrier signals, which makes it possible to delegate the energy expensive
carrier generation to mains-powered devices like WiFi routers or ZigBee hubs~(see Section~\ref{subsec:carrier_generation}).

%TODO This paragraph needs clarification
Keeping the carrier and backscatter signal separated in frequency, also introduces a new challenge in
the design of the  tag. Traditional CRFID tags only modulate the carrier with information while
our architecture requires the carrier to be frequency-shifted  and modulated.  Recent low-power
designs of such tags are implemented on ASICs 
and are in simulations~\cite{kellogg2016passive,interscatter,fmbackscatter}, or designs
built using off-the-shelf components modulate ambient signals to amplitude modulated signal~\cite{zhang_enabling_2016}.
We present a backscatter tag  that can shift and frequency 
modulate the carrier signal. The tag consumes \SI{70}{\micro\watt} 
and \SI{650}{\micro\watt} while operating at \SI{868}{\mega\hertz}
and \SI{2.4}{\giga\hertz} respectively.

In using commodity devices as carrier generators, our architecture operates in the
shared  \SI{2.4}{\giga\hertz} ISM band and
encounters the problem of cross-technology interference~(CTI). 
To mitigate the harmful effects of CTI, we demonstrate two mechanisms: First, we show  that
leveraging multiple wireless devices to generate carrier signals
at different  frequencies can enable simultaneous backscatter transmissions. 
When coupled with several receivers the probability of reception
improves, even under CTI. Second, our results demonstrate that by changing the frequency of
the carrier signal, we can make backscatter transmissions avoid CTI.

Note that in our use cases, as well as in the rest of the paper,
we focus on the uplink from the backscatter tag 
to the reader since most sensing
applications are constrained by this link~\cite{BackFi, HitchHike}.
We can support  receptions using existing low-power receiver
designs~\cite{wifibackscatter}. Existing 
CRFID readers combine energy delivery with communication on the same RF carrier,
which has been shown to be inefficient~\cite{zhang2014enabling,gummeson2010limits}. 
We hence decouple the RF energy delivery from the reader. Our backscatter tag 
still consumes \SI{}{\micro\watt s}, which 
can be easily provided by many ambient energy sources~\cite{naveddtosn}. 
Further, LoRea, if needed, can support RF-based 
energy harvesting by using the harvester design 
presented by Talla et al.~\cite{talla2015powering}.

The paper proceeds as follows. We discuss background and related 
work in Section~\ref{sec:background}. Next, we discuss the design, 
implementation and cost analysis of our architecture 
in Section~\ref{design}. In Section~\ref{evaluation} we present our 
experimental evaluation. Section~\ref{applications} discusses two 
challenging applications our architecture enables.
Before concluding,
we discuss some issues related 
to our architecture in Section~\ref{discussion}.

\section{Background and State of the Art}
\label{sec:background}
This section presents a background on backscatter and self-interference as well as work related to \system.

\subsection{Backscatter primer}
\fakepar{Overview}
\label{sec:background:backscatter}
When radio frequency (RF) signals interact with an antenna, they are
absorbed or reflected by a varying amount dictated by the antenna's radar cross section~(RCS).
Backscatter devices control the RCS by changing the 
impedance of the circuit connected to the antenna,
switching the antenna to either reflecting or absorbing mode.
This mode change induces
minute variations in the ambient signal which can be observed by an RF receiver.

Consider an RF emitter that is transmitting a signal $S_{rt}(t)$ that
reaches the antenna of the backscatter device. The device selectively
reflects or absorbs $S_{rt}(t)$.
%If a receiver observes the reflected
%signal coming from the  backscatter device, the signal received
At a receiver, the reflected signal 
$R(t)$ consists of two components: $S_{rt}(t)$ coming directly from the emitter device 
and $S_{bt}(t)$
caused by the minute variations induced by the backscatter operation. The resulting
signal  can be expressed as: 
\begin{equation}
	R(t)=S_{rt}(t) + \sigma B(t)S_{bt}(t) \label{eq:receiver_signal}
\end{equation}

In the above equation, $\sigma$ is the RCS of the device, and $B(t)$ is 
the bit sequence transmitted by the device, that is $1$ when reflecting,
and $0$ when absorbing.  In traditional RFID readers, the reader both generates the
carrier signal and receives the backscattered signal. The reader generates a 
tone signal or a pure sinusoid at frequency $f_{c}$, and the
backscatter tag reflects at the same frequency, i.e., in the above
equation both components are at the same frequency.

\fakepar{Backscatter as mixing process}
%As we had seen  in  the
Equation~(\ref{eq:receiver_signal}) shows that
the signal backscattered from the tag is proportional to 
the product of the baseband signal $B(t)$ generated by the tag
and the ambient signal $S_{bt}(t)$ at the tag.
If we assume that backscatter
readers generate a carrier signal at a specific  
frequency $f_c$, while the tag is changing the RCS
of the antenna at a frequency of  $\Delta f$, the resulting
signal (product $\sigma B(t)S_{bt}(t)$ in Equation \ref{eq:receiver_signal}) can be expanded to the following form: 
\begin{equation}
	 2\sin(f_c t)\sin(\Delta f t) = \cos[(f_c + \Delta f)t] -
\cos[(f_c - \Delta f)t]. \label{eq:mixing_equation}
\end{equation}

The result is that the backscattered signal appears at an offset
$\Delta f$ on the positive and the negative sides
of $f_c$, the centre of the carrier signal.
This displacement  helps the backscatter tag both
modulate the carrier and reduce interference from the carrier to the weak backscattered
reflection~\cite{kellogg2016passive,HitchHike}.

\subsection{Self-interference mitigation}

Self-interference in wireless systems occurs when a radio transmits and receives simultaneously at the same
frequency. This makes it a problem particularly for full-duplex
radios~\cite{Jain:2011:PRF:2030613.2030647,Choi:2010:ASC:1859995.1859997,Bharadia:2014:FDM:2616448.2616482},
where the strong transmitted signal can overwhelm the sensitive
receiver.  RFID readers are full-duplex in the sense that they
must receive the weak backscattered signals while transmitting the
unmodulated carrier in the same frequency, and thus suffer from the same
issue.

The problem is exacerbated in RFID systems because the reader, when
querying the tags for their IDs, must also
provide energy to the (passive) tags and hence transmits a powerful
carrier signal~(usually \SI{30}{dBm}). 
To mitigate self-interference, RFID readers typically 
employ sophisticated mechanisms to recover the weak backscattered signal.
These mechanisms are usually a combination of methods that isolate the
carrier using circulators, employ RF cancellation to attenuate the
carrier signal on the receive chain and finally separate the interfering
carrier signal from backscatter transmissions~\cite{braidio}.  These
methods increase the power consumption, complexity and cost of the
reader.  For example, Impinj's R2000 RFID chip consumes an additional
500~mW when its self-interference cancellation circuit is
enabled~\cite{impinjr2000}.  Furthermore, the use of circulators comes
with an insertion loss penalty that reduces the signal strength of the
received signal, which in turn limits the achievable range.  CRFID
applications typically employ conventional RFID readers to receive
backscattered transmissions.

SDR-based readers are also often used to query CRFID tags.
These readers do not include any specialized hardware to
reduce self-interference from the strong carrier. Instead, they resort to
operating in a bistatic configuration~\cite{buettner_software_2011} but nevertheless
the achievable range is reduced to only a few meters~\cite{hu2015laissez,zhang2014ekhonet,BuZZ}. 

Liu et al. present a design to reduce self-interference and enable
full-duplex operation on ambient backscatter devices~\cite{Liu:2014:EIF:2639108.2639136}. 
Their design achieves a range of only a few meters. %\tv{plural (these). Do you refer to Liu only or also all the above?}
Recent backscatter systems leverage the spectral mixing property
of backscatter transmissions to 
shift the frequency of backscatter transmissions away from the carrier to
reduce
%self-interference~\cite{kellogg2016passive,HitchHike,blebackscatter,passivesensortags,passivezigbee,zhang_enabling_2016}.
self-interference~\cite{kellogg2016passive,HitchHike,blebackscatter,passivezigbee,zhang_enabling_2016,vlcs}.
We build upon these designs to develop an  inexpensive reader
for CRFID devices that achieves a high communication range.

Ma et al.~\cite{ma16loc} use non-linear elements
attached to the WISP platform to reduce self-interference
and achieve accurate 3D localization.
While Ma et al.~\cite{ma16loc} also reduce self-interference, 
we target a different problem: by shifting
the carrier we reduce self-interference to lower the cost of
the backscatter reader and achieve the highest demonstrated 
range with backscatter systems.

\subsection{Low-power readers}

There have been prior attempts to develop low-cost backscatter
readers. Braidio is a backscatter reader that can switch between
active and passive radios depending on the energy constraints of the
host device~\cite{braidio}.  Similar to our architecture, Braidio can
function as a low-cost and low-power backscatter reader, but achieves
a maximum range of \SI{1.8}{\metre} at \SI{100}{kbps}.  As a comparison, we
achieve a significantly higher range due to three primary reasons: First,
Braidio uses passive receivers similar to the ones found on RFID tags
resulting in a sensitivity of approximately \SI{-60}{dBm}. By
contrast, the receivers employed in our architecture are narrow-band
radios with a high sensitivity level~(\SI{-124}{dBm}). Second, we separate the weak
backscattered signal from the strong carrier which improves the SNR of
the backscattered transmissions.  Finally, we use a bistatic
operation of the reader which further reduces self-interference.
Another notable attempt is from Nikitin~et~al., who design a simple
low-cost reader~\cite{nikitin2013simple} but achieve a range of only
\SI{15}{cm}.

\fakeparnodot{Do we still need backscatter readers?} 
Recent systems demonstrate the 
ability to synthesise transmissions compatible
with WiFi (802.11b)~\cite{kellogg2016passive}, BLE~\cite{blebackscatter} 
and ZigBee~\cite{interscatter,passivezigbee} while consuming \SI{}{\micro\watt s} 
of power ameliorating the need for a separate reader device. 
These state-of-the-art systems operate in bistatic mode with the tag co-located
with the carrier generator, and demonstrate a  range of 
tens of meters. For example, Passive WiFi~\cite{kellogg2016passive} shows that WiFi transmissions
synthesised using backscatter communication can be received up to a distance of \SI{30}{\meter}
with the tag located \SI{1}{\meter} from the carrier generator. Our architecture, under a similar setup and frequency,
achieves a range of over \SI{200}{\meter}. While the ability to communicate over tens of meters with \SI{}{\micro\watt s} of
power consumption enables
novel applications such as connected implants~\cite{interscatter}, 
it is not sufficient for many applications that require even
longer communication range.

%One of the primary reason for the low range achieved is the fact that modern wireless standards 
%operate at bitrates in the order of Mbps,  which has the effect of reducing the sensitivity levels of the
%radio. For example, a typical sensitivity level on modern WiFi transceivers~\cite{CC3200} operating at \SI{54}{MBps} 
%is \SI{-74}{dBm} as compared to a sensitivity level of \SI{-124}{dBm} at a bitrate of  
%\SI{625}{Bps} on an IEEE 802.15.4g transceiver~\cite{CC1310}. Backscatter reflections on the other hand are usually weak, which together
%with the poor sensitivity levels of commodity radio transceivers leads to a low communication range.

\subsection{Ambient backscatter systems}

Ambient backscatter leverages 
radio signals such as TV
transmissions~\cite{ambientbackscatter} or WiFi
traffic~\cite{BackFi,wifibackscatter, HitchHike}
to dispense with the need for an external reader
or a device to generate the external carrier. 
Parks et al. demonstrate passive tag-to-tag 
communication using ambient TV
signals~\cite{ambientbackscatter}. 
They further improve on the design 
to enable through-the-wall 
operation and achieve high 
throughput~\cite{turbochargebackscatter}.
Ambient backscatter using TV signals, however, is limited 
to operate only in the vicinity of TV towers where the signal 
is strong enough~(approx. \SI{-30}{dBm}) 
with a limited range of \SI{30}{\metre}~\cite{turbochargebackscatter}.

On the other hand, some recent systems backscatter ambient WiFi signals. Kellogg et al.
demonstrate the feasibility of backscattering WiFi signals and
receiving on commodity smart phones~\cite{wifibackscatter}
at a short range \SI{2.1}{m}.
Zhang et al. improve upon WiFi backscatter and achieve a range of
\SI{4.8}{\meter} by using frequency-shifting to  reduce interference
from WiFi transmissions to weak backscatter signals.
Bharadia et al. demonstrate high-throughput WiFi backscatter
to distances up to \SI{5}{m}~\cite{BackFi}.  
Their design uses extensive self-interference 
cancellation techniques  at the receiver which
makes the system both complex and expensive. 
HitchHike~\cite{HitchHike} enables communication with
commodity WiFi radios by changing the codewords of WiFi signals
and achieves a range of \SI{54}{m}.  WiFi backscatter systems do not
require a dedicated carrier generating device. However,
these systems occupy a significant portion of the 
license free spectrum due to the large bandwidth~(\SI{22}{\mega\hertz}) of 
WiFi signals. As a comparison our architecture
achieves a significantly higher range, and uses the spectrum 
efficiently due to narrow-bandwidth 
transmissions.

\section{Design}
\label{design}
In this section, we present our architecture, the design of the backscatter reader,
the tag, the mechanisms to bring frequency diversity to
backscatter tags and a cost analysis of the architecture. 

\subsection{Architecture}

Our architecture is depicted in Figure~\ref{fig:architecture}.
In contrast to traditional RFID
readers, the reader is split into one or more carrier generators and
one or more receivers. Part of our architecture is a tag that shifts and modulates
the carrier signals when backscattering it.
The rest of this section describes these components.

\subsection{Reader} 
\subsubsection{Decoupling in Frequency and Space}\hspace*{\fill} 

%In our architecture, the reader achieves significantly higher receive sensitivity 
%and an order of magnitude lower cost as compared to state-of-the-art
%CRFID readers primarily due to the way we tackle self-interference.
%Reducing interference from the carrier signal is important since otherwise
%the strong carrier overwhelms the backscattered signal
%which is inherently weak~\cite{braidio}.  We next describe how our architecture
%achieves this.

As described in Section~\ref{sec:background}, 
tackling self-interference is important when aiming for low cost and high range.
Our architecture achieves this by decoupling in frequency and
space:

We keep the carrier signal and the backscattered signal on \emph{different frequencies}. 
As opposed to conventional readers, where the carrier signal and
the backscattered signal overlap in frequency, we 
deliberately place the carrier an offset $\Delta f$ away from the frequency on which the reader 
listens. Modern
radio transceivers can
greatly attenuate signals present in the adjacent bands. For example, the CC2500 attenuates 
a signal present \SI{2}{\mega \hertz}
away from the tuned frequency by almost \SI{50}{\deci \bel} (Figure~\ref{selfintfcancel}).
This separation between backscatter signal and the carrier
significantly attenuates the carrier signal without using the complex techniques and
components employed on existing readers. 

%\emph{Second},  existing RFID and CRFID tags backscatter on baseband frequencies~\cite{wisp, ma16loc}. 
%In our architecture the tags \emph{backscatter
%on an intermediate frequency}.
%We leverage the spectral mixing process inherent to
%backscatter transmissions to both modulate the
%carrier, and also to
%shift the modulated signal to the desired 
%frequency $f_r = f_c + \Delta f$ on which the reader listens for transmissions. 
%This helps to improve the SNR of the signal, and to also receive information,  
%while the carrier is located some offset away.
%\tv{to me the difference between First and Second is not really clear}
%\cp{Second that}

Our architecture also \emph{spatially decouples carrier
generation from reception}.
Spatial separation further reduces interference at the receiver from the carrier signal due to propagation loss~\cite{buettner_software_2011}. 
On existing RFID readers the carrier generator
and the receivers are usually co-located, hence have to employ complex components
like circulators to reduce self-interference. Our decoupled architecture also enables us
to reduce the power consumption of the receiver~(see Section~\ref{parkinglotexp}). 
Furthermore, when operating in the \SI{2.4}{\giga\hertz}
band, our architecture can leverage existing devices (e.g. WiFi access points or
sensor nodes) as carrier generators. 
Using commodity devices that are part of the infrastructure as carrier generators helps 
improve the scalability of the system.

\subsubsection{Receiver}\hspace*{\fill} 

\label{sec:receiver}
%The receiver %is the  heart of our architecture since it
%is responsible for receiving backscattered transmissions from the tags. 
Designing the
reader from scratch opens the design space
to select the transceiver and important parameters like
intermediate frequency, bandwidth and the modulation scheme.

\fakepar{Transceiver}
We select commodity narrowband 
transceivers to receive backscatter transmissions. 
Such transceivers present two major advantages:
\emph{First},  they are highly
configurable in that we can select both
modulation scheme and bitrate.
This enables us to significantly
reduce the bitrate. Since the receive sensitivity
improves drastically at lower bandwidth, we can therefore
significantly extend the communication range.
\emph{Second}, supporting only basic link-layer functionality,
without support for high layer protocol stacks
like BLE or WiFi, enables maximum configurability
and a clean slate-design of the reader.
Most sensing applications
send only small amounts of data~\cite{oppermann2014decade}. While
these applications can benefit from high bitrates, a low bitrate is not detrimental to the application's performance.  To support high bitrates,
we can also operate the reader at high bitrates with reduced sensitivity.

In our implementation
we select the Texas Instruments CC2500~\cite{CC2500} radio transceiver for the \SI{2.4}{\giga\hertz}
ISM band, and the CC1310~\cite{CC1310} for the \SI{868}{\mega\hertz} band
because of their superior 
configurability, low-power and narrow bandwidth receptions.

\fakepar{Intermediate frequency selection} 
We use spatial and frequency separation to reduce interference from the carrier signal. The intermediate frequency $\Delta f$ for the frequency separation has to be large enough to significantly attenuate the carrier signal, leveraging the transceiver's adjacent channel rejection; but as small as possible because the tag's power consumption increases with $\Delta f$~\cite{zhang_enabling_2016}. 

%, corresponds to 
%the minimum frequency needed to significantly attenuate the carrier signal, leveraging
%the adjacent channel rejection ability of the transceiver.
%the transceiver's ability to reject adjacent channels.
%As power consumption
%increases with frequency, the selection of $\Delta f$ is crucial to achieve low power consumption at the tag.

\begin{figure}
    
    \subfigure[CC2500 (\SI{2.4}{\giga\hertz )}]{
        \includegraphics[width=0.9\columnwidth]{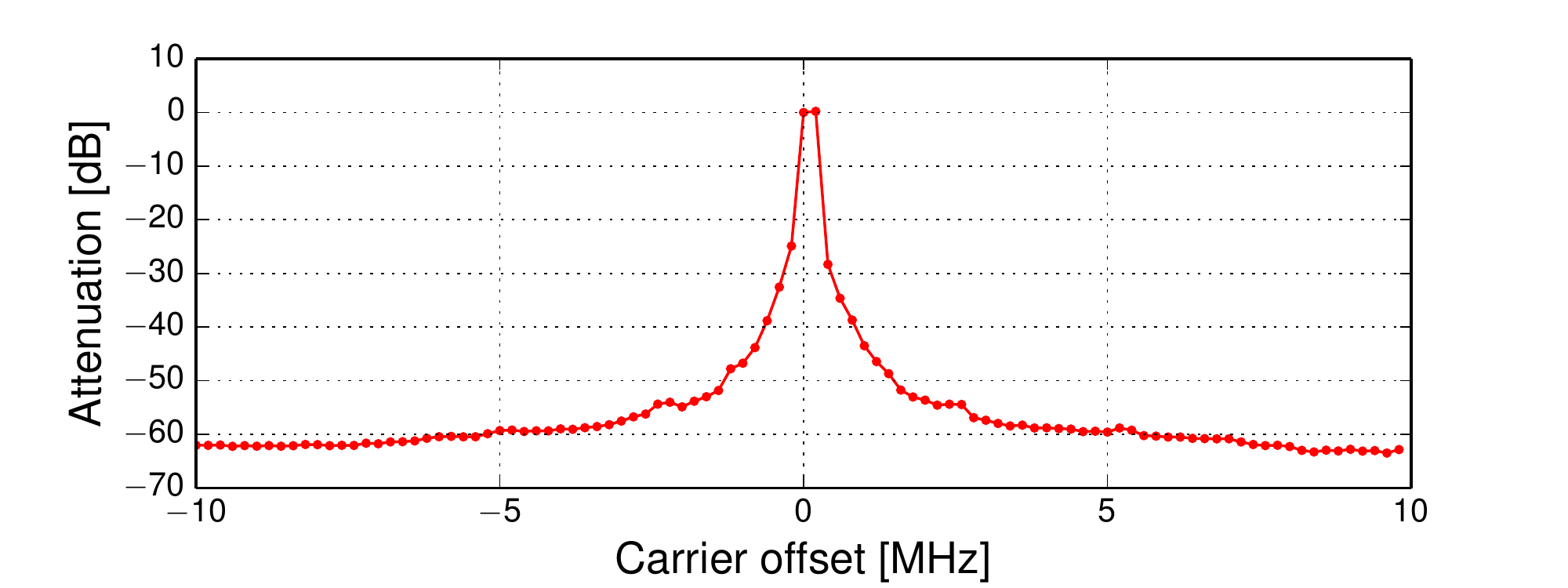}
        \label{fig:cc2500intfrej}
    }
            \vspace{-4mm}

    \subfigure[CC1310 ( \SI{868}{\mega\hertz} )]{
        \includegraphics[width=0.9\columnwidth]{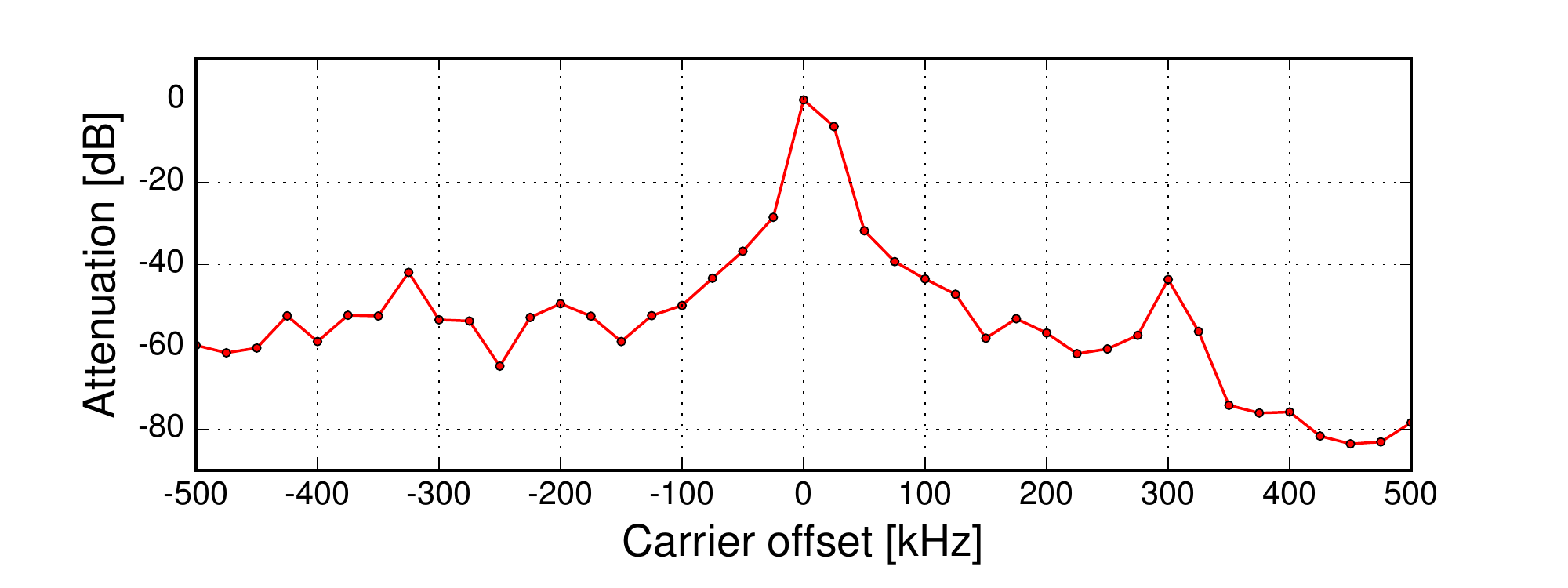}
        \label{fig:cc1310intfrej}
    }
        \vspace{-4mm}

    \caption{\capt{Carrier interference rejection.}
				 The transceivers reduce interference from the carrier located \SI{2}{\mega \hertz} (\SI{2.4}{\giga \hertz}) and \SI{100}{\kilo \hertz} (\SI{868}{\mega \hertz}) away by more than \SI{50}{\deci \bel}.}
    \vspace{-4mm}
\label{selfintfcancel}
\end{figure}

The choice of $\Delta f$ is transceiver-dependent.
We conduct experiments to determine $\Delta f$ for the transceivers we
use.
%for the \SI{868}{\mega\hertz} and \SI{2.4}{\giga\hertz} 
We set up a software defined radio~(SDR) %, and position its antenna about
%\SI{30}{\centi\metre} away from the transceiver's antenna.  We program
%the SDR
to perform a frequency sweep over a \SI{20}{\mega\hertz}
(\SI{2.4}{\giga\hertz}) and \SI{2}{\mega\hertz} (\SI{868}{\mega\hertz})
range centered on the receiver's tuned frequency $f_c$.
Meanwhile, the receiver records the received signal strength at the
different carrier offsets.
%We program the SDR to generate an
%unmodulated carrier signal with the SDR's maximum output power of \SI{12}{dBm}.
%We tune its carrier frequency to the receiver's
%centre frequency $f_c$.  Furthermore, we program the receiver to 
%measure the noise floor repeatedly.
%When the carrier is at the receiver's centre frequency,
%the noise floor is significantly high because of significant interference
%from the carrier signal.  Next, we program the SDR to slowly sweep the 
%frequency of the carrier signal, until it 
%reaches a frequency \SI{10}{\mega\hertz} (\SI{2.4}{\giga\hertz})
%and \SI{1}{\mega\hertz} (\SI{868}{\mega\hertz}) on  
%both sides of the centre frequency $f_c$. We find the mean noise floor
%at different offsets of the carrier, and the change in the noise floor from the levels
%when the carrier was present at $f_c$.  
Figure~\ref{selfintfcancel} depicts the result normalized to the minimum
rejection which naturally occurs at zero offset.
The carrier rejection improves by almost \SI{50}{\deci\bel} when the
carrier is shifted \SI{2}{\mega\hertz} away from $f_{c}$ for the CC2500.
The rejection improves by \SI{50}{\deci\bel}  when the carrier is  \SI{100}{\kilo\hertz}
away from $f_{c}$ on the CC1310, without much further improvement after
that.
Based on these results, we consider $\Delta f = $
\SI{2}{\mega\hertz} and \SI{100}{\kilo\hertz} 
as a good trade-off for the %implementation with the
two transceivers.

\fakepar{Selecting Modulation Scheme}
Since we redesign the reader from scratch, we can select
the modulation scheme. The transceivers in our architecture
support both On-Off Keying (OOK) and Frequency-Shift Keying (FSK).  

Existing CRFID tags usually employ amplitude modulation for communication,
as the passive receivers employed on these tags are often limited to
amplitude demodulation using simple envelope detectors~\cite{ambientbackscatter}.
We choose FSK since it provides several advantages: \emph{First}, FSK
is a constant-envelope modulation~\cite{rappaport_wireless_2002} and 
offers 
robustness against fading. \emph{Second}, FSK is more robust to noise 
than amplitude modulation since it can 
achieve a lower Bit Error Rate~(BER) for the
same signal-to-noise 
ratio~\cite{kimionis_increased_2014,rappaport_wireless_2002}. 
We employ 
a frequency deviation of \SI{13}{\kilo\hertz} and \SI{190}{\kilo\hertz} 
between the bit 0 and 1 for the CC1310 and the CC2500 respectively.

%\begin{table*}[]
%\centering
%\scriptsize
%\begin{tabular}{lrrrrrrcr}
%\toprule
%Component & CC2420 & CC3200 & CC2500 & Beaglebone & MSP430 & Frontend & RFID Reader\\
%\midrule
%Task & \multicolumn{2}{c}{Carrier generator}  & Receiver & \multicolumn{3}{c}{Backscatter tag} & RFID Reader\\  

%Power (mW) & 52  & 687 & 39.9-58.8 & 10.9 & 7.2 & 0.0003  & 11500-15000\\
%Cost (\$) & 11.1 & 29 & 5.0 & 45 & 16 & 1  & 1585.0 \\
%\bottomrule
%\end{tabular}%
%\caption{Power consumption and cost of \system's components in comparison with RFID Reader}
%\label{table:pc}
%\end{table*}

\subsubsection{Carrier Generation}\hspace*{\fill} 

\label{subsec:carrier_generation}

% vim: textwidth=72 spell spelllang=en_us
A crucial task of our architecture is the generation of the 
carrier signals that are then reflected by the tags.
Traditional readers delegate this task 
to a single device.  Instead, our architecture uses a
bi-static configuration and spatially
separates the carrier generation and the reception. We describe this next.

%TODO We should add a reference to figure 3 in this section
\fakepar{Monostatic vs. bistatic setup} Most
existing backscatter systems follow a \textit{monostatic}
setup, in which the RFID reader uses the same antenna for emitting a
suitable carrier and for receiving the transmissions from the backscatter
tags~\cite{kimionis_increased_2014,nikitin2008antennas}. An advantage with this setup is its
conceptual simplicity. However, as discussed in
Section~\ref{sec:background},
monostatic setups require the
reader to perform complex interference cancellation which increases
the cost and complexity.% because the
%emitted carrier is much stronger than the backscatter transmissions from the 
%tags.

\begin{figure}
    
    \subfigure[Signal strength in monostatic setup, as the
    tag's distance to the reader increases.]{
        \includegraphics[width=0.9\columnwidth]{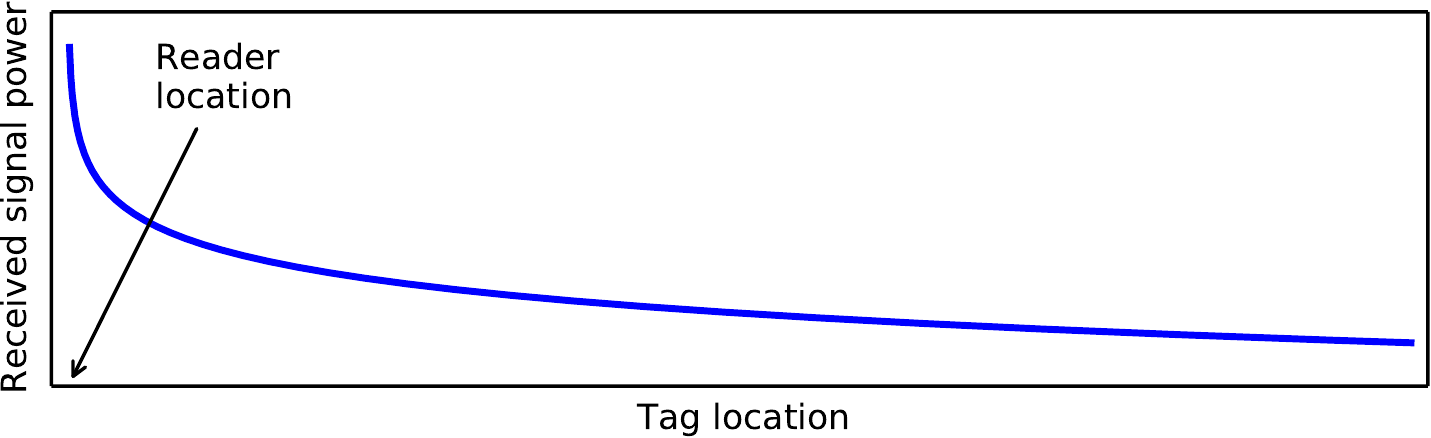}
        \label{fig:design:carrier:mono}
    }
    \subfigure[Signal strength in bistatic setup, as the tag is placed
    on a straight line between receiver and carrier generator.]{
       \includegraphics[width=0.9\columnwidth]{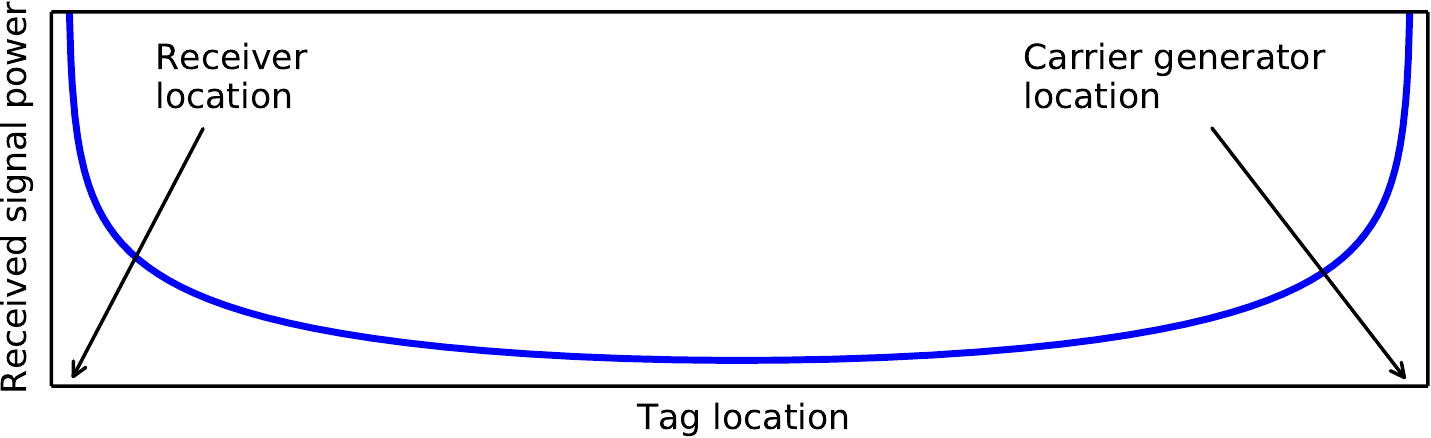}
        \label{fig:design:carrier:bi}
    }
    	\vspace{-4mm}

  \caption{\capt{Signal strength for monostatic and bistatic setups.}
        A bistatic setup increases the range, providing more
        locations from which the backscattered signal can be
    	received with a high signal strength.}
    	\vspace{-4mm}
    	\label{monostatic}
\end{figure}

Monostatic configurations also limit the communication range.  
Consider the strength $P_r$ of a backscattered signal at the reader in free
space~\cite{kellogg2016passive,nikitin2008antennas}, given by
\begin{equation*} P_r = \left ( \frac{P_tG_t}{4\pi d_1^2} \right ) K
\left ( \frac{\lambda ^2   G_r}{4\pi d_2^2 4 \pi} \right ).
\end{equation*}
Here, $\lambda$ is the carrier's wavelength, $P_t$ is the power of the
carrier, and the factor $K$ accounts for the return loss and antenna
gains at the backscatter tag.  $G_t$ and $G_r$ represent the antenna
gain for transmitting the carrier and receiving the backscattered
signal, respectively.  Similarly, $d_1$ denotes the distance of the
backscatter tag to the carrier generator and $d_2$ denotes the distance
of the tag to the receiver. Thus, in a monostatic configuration, $d_1 =
d_2$ and $G_t = G_r$.    As expected, minimizing the distance to the RFID reader
maximizes the received signal strength.

In contrast, our architecture uses a \textit{bistatic} configuration, in which
receiver and carrier generator do not share the same antenna and can be
spatially separated. This means that for our architecture $d_1$ does not need to
be identical to $d_2$. An interesting property of the bistatic configuration resulting
from the duality of $d_1$ and $d_2$ is that the received signal strength is high 
if the backscatter tag is located in proximity to either the receiver or the carrier
generator, as we illustrate in the Figure~\ref{monostatic}.

Another advantage of the bistatic configuration is that the interference
from the carrier can also be reduced due to path-loss
provided that carrier generator and receiver are separated in space~\cite{buettner_software_2011}. This
further reduces the cost and complexity of the reader.

Finally, generating the carrier signal is one of the most energy consuming
tasks on the reader.  Co-locating the carrier generator together
with reception circuitry results in a significant increase in the power
consumption of the device, which makes it difficult to
operate in mobile scenarios.  The bistatic setup also helps achieve such 
capability (see Section~\ref{mobilereader}).

\fakepar{Generating carriers} We are generally surrounded by 
commodity devices equipped with WiFi, BLE or ZigBee radios.  Leveraging
these devices to generate the carrier signal could
significantly improve the scalability of our system. 
Interscatter~\cite{interscatter} demonstrated that sending a special payload could help to 
generate short carrier signals from BLE radios.  While
BLE radios are very common, they are mostly found on smartphones 
or fitness trackers which are usually
battery-powered. Delegating the energy-expensive carrier generation to 
devices operating on batteries might be detrimental to their life time. 
On the other hand, WiFi access points and ZigBee 
hubs are ubiquitous and are usually mains-powered, 
making them suitable to generate carrier signals.

To use WiFi or 802.15.4 devices to generate the carrier signal,
we take advantage of the fact that most radio transceivers
provide access to a special test mode that generates 
an unmodulated carrier signal. The radios provide 
%easy access to 
this test mode 
to enable regulatory compliance testing. We leverage this
mode to generate carrier signals from WiFi and 802.15.4
radio transceivers. In Section~\ref{sec:eval}, we use TelosB sensor nodes~\cite{polastre2005telos} that
feature a CC2420 radio chip~\cite{CC2420}, and the WiFi radio CC3200~\cite{CC3200} to generate an
unmodulated carrier signal.  Our architecture can also take advantage of
the
carrier signals generated by Interscatter on BLE radios.
Since a carrier wave does not contain any
information, the generation of carriers does not need to be coordinated in a
deployment.  Indeed, \system can use any combination of carrier
generators as we show in Section~\ref{sec:freqdiversity}.
%TODO Note that carrier do not carry information but they can indeed jam other information-carrying signals. This might be a reason to require coordination of carrier generation

\fakepar{Carrier frequency} Apart from using the sub \SI{}{\giga\hertz} frequency band
that  conventional CRFID systems use, we primarily operate in the \SI{2.4}{\giga\hertz} ISM band. A key motivation
for this decision is the uniform world-wide availability of the \SI{2.4}{\giga\hertz} band and
its relatively high permissible transmit power. Furthermore, at \SI{2.4}{\giga\hertz} 
our architecture can also leverage existing deployed
devices like WiFi radios and sensor nodes to provide a carrier signal.

\fakepar{Power consumption} The CC3200 WiFi radio 
consumes \SI{687}{\milli\watt}
and the CC2420 802.15.4  radio consumes \SI{54}{\milli\watt}
%of power
to generate the carrier signal.
The high power consumption required to generate the carrier signal is 
common to all backscatter systems~\cite{kellogg2016passive,interscatter}.
However, our architecture ameliorates the particular issue by enabling
externally powered devices such as WiFi routers or ZigBee hubs to act as carrier generator.

\subsection{Backscatter tag}
% vim: textwidth=72 spell spelllang=en_us

%TODO In all rigour, all those works usually talk about a custom IC implementation and not about ASICs. We should probably talk about that. It would make this argument even stronger.
\fakepar{Design philosophy}  Existing systems, like
Interscatter~\cite{interscatter}, Passive WiFi~\cite{kellogg2016passive} or FM backscatter~\cite{fmbackscatter} present
an IC design of the  
backscatter tag in a simulated environment, while the actual experiments were conducted with
prototypes
built using FPGAs or function generators
that have a power consumption similar to low-power
radios.  Fabricating ICs especially 
in small quantities is prohibitively expensive. 
Our key design philosophy is to use only off-the-shelf
components in the design of backscatter tag which  consumes \SI{}{\micro\watt s}
of power. This brings the ultra-low power 
designs of backscatter tag to the wider research community immediately.

%\fakepar{Failed attempts} Our initial attempt to design backscatter tags 
%used commodity MCUs~(MSP430) to both generate the intermediate frequency signal
%and to modulate the signal with information. However, the use of MCU increased
%the power consumption to a few \SI{}{\milli\watt s}. 
%Next we investigated a design presented by Zhang et al. that uses ring-oscillators
%to generate an intermediate frequency signal at very low power consumption~\cite{zhang_enabling_2016}. Their design, however,
%is not applicable to our architecture: Firstly because of the much higher frequency variability of the ring-oscillators
%as compared to the tolerance region of our receivers, and secondly the
%mismatch in modulation schemes (OOK vs FSK).

% We also investigated schmitt-trigger
%oscillators. However, owing to
%asymmetric duty cycle (33\%) and high power consumption in non-linear regions of the oscillator, 
%we decided against using them in our tag.

\begin{figure}

    \subfigure[Schematic]{
        \includegraphics[width=0.4\linewidth]{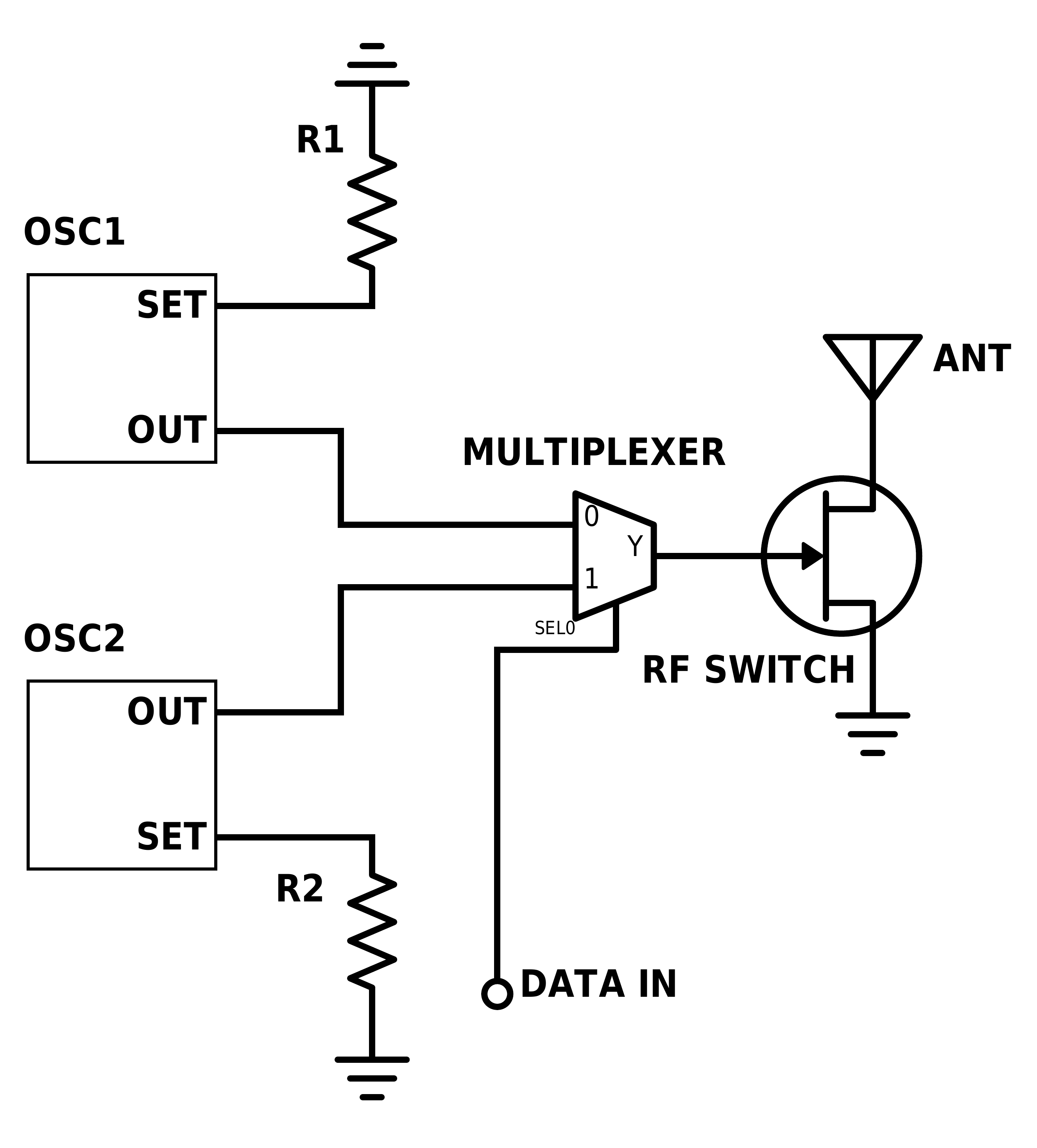}
        \label{backscattertag}
    }
        \subfigure[Prototype]{
        \includegraphics[width=0.55\linewidth]{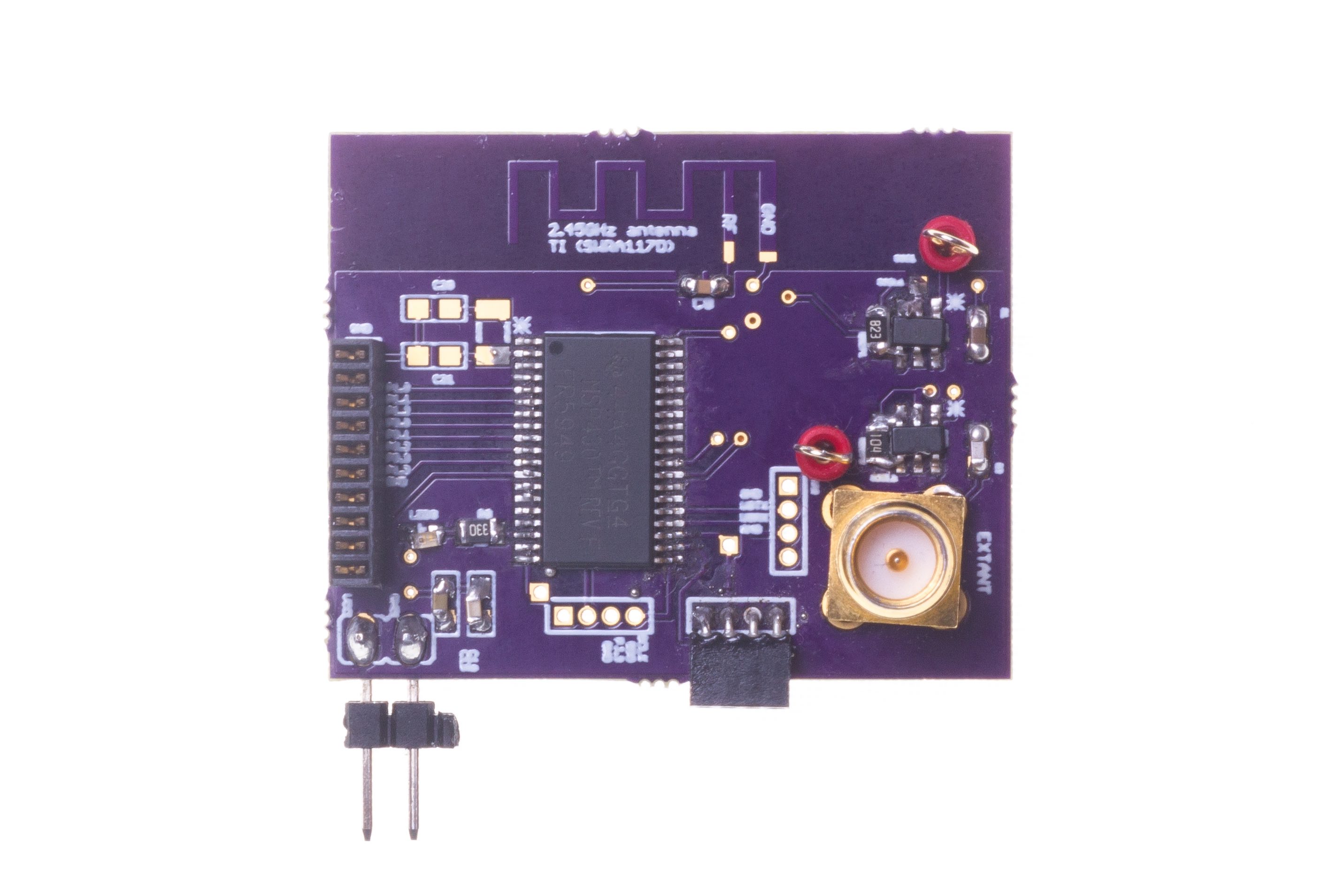}
        \label{backscatterhardware}
    }
    
    \caption{\capt{Backscatter tag schematic and prototype.}
				 The tag shifts and
				modulates an ambient carrier with microwatts of power. }

\label{hardwaretag}
\end{figure}

\fakepar{Backscatter tag design} We design  our tag on a
two-layer FR4 PCB. We present a simplified schematic
of the tag in Figure~\ref{backscattertag}. 
At a high level our tag works as follows: First,
using two oscillators we generate digital signals
corresponding to the two frequencies (0 and 1) 
of the FSK signal.
Next,  the tag selects one of the two 
signals using a multiplexer chip based on the
information it wants to send. Finally, the resulting
signal is used to control an RF switch, which switches the
antenna to reflecting or absorbing state modulating
the ambient signal with the information to transmit. 
We show the hardware prototype of the tag  
in the Figure~\ref{hardwaretag}.

In our design, the Analog Devices HMC190BMS8 is the RF
switch~\cite{hmc190}.
This switch has also been employed in recent 
backscatter systems~\cite{interscatter,kellogg2016passive}.  
We select the Linear technology LTC6906 and the LTC6907 
oscillators for the \SI{868}{\mega\hertz} and \SI{2.4}{\giga\hertz} 
tags respectively due to their ultra-low power consumption. As multiplexer, we use
Analog Devices ADG904. 

We have measured the return loss of our  tag as 3~dB, which is similar
to recent designs~\cite{interscatter,kellogg2016passive}.
Backscatter transmissions have a side effect of creating an undesired mirror signal (see Eq.~\ref{eq:mixing_equation}).
Our present font-end does not remove this image. In the future, we will
incorporate the
design presented by Zhang et al.~\cite{HitchHike} to resolve this which might further
improve the range. However, in-spite of the undesired image, owing to narrow-band
transmissions, the backscatter signal, undesired mirror image and  the carrier signal
occupy less than \SI{4}{\mega\hertz} of bandwidth  at \SI{2.4}{\giga\hertz}  which is less than
the channel spacing of 802.15.4 which eases the coexistence with other wireless networks.

For faster prototyping, we also develop a tag based
on the Beaglebone~Black embedded platform~\cite{beaglebone}~($\sim$ \$45)
and the MSP430FR5969
MCU that are also used on present CRFID platforms.

\begin{figure}
 
        \includegraphics[width=\columnwidth]{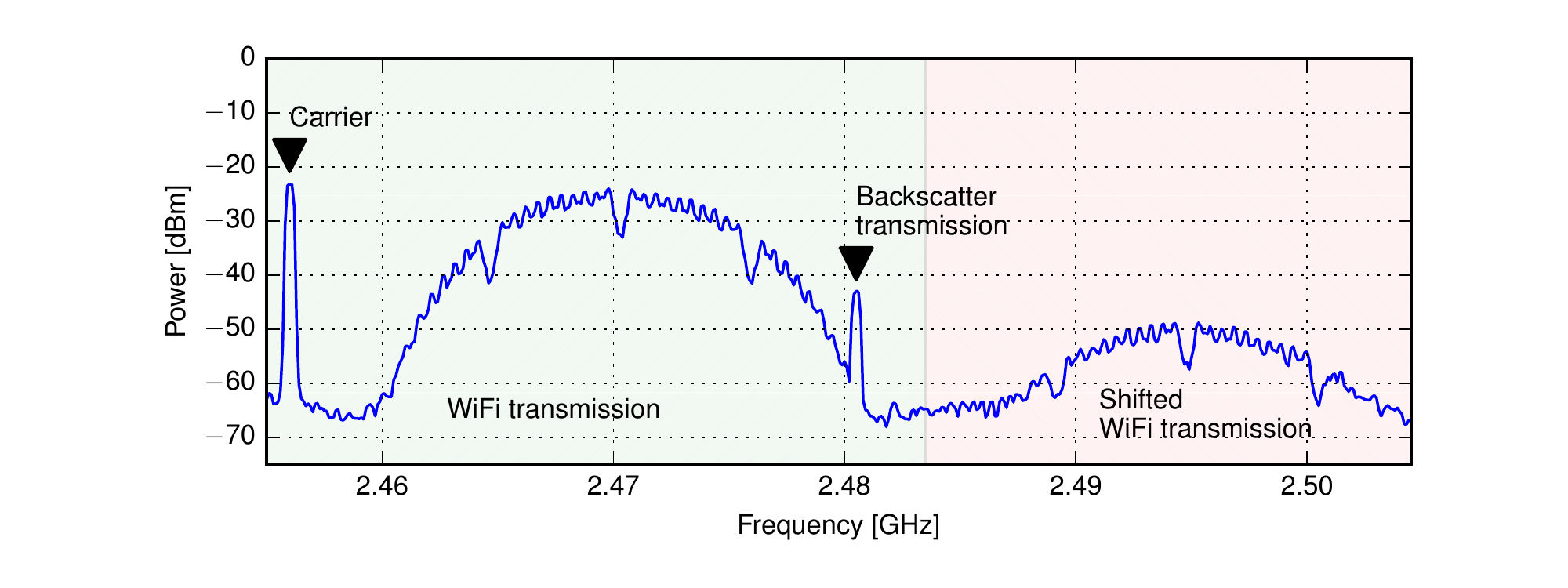}
    \caption{\capt{Large frequency deviations at backscatter tag. } A large intermediate frequency at the backscatter
    tag also shifts ambient WiFi transmissions out of the license-free
				ISM band. The red shaded (\textbackslash\textbackslash) region
    is outside the unregulated band.}
        \label{fig:spectrumshift}
\end{figure}

\fakepar{Power consumption}  The power consumption of the backscatter tag
is dependent on both the intermediate frequency  at the tag,
and the operating voltage. As power consumption decreases with operating
voltage~\cite{zhang_enabling_2016}, we operate the backscatter tag at the lowest operating voltage, which we
found to be \SI{2.1}{\volt}, the minimum required for the oscillators.
To measure power consumption, we use a highly sensitive Fluke 289 multimeter connected 
in series with the backscatter tag. Table~\ref{table:comp} shows the results of these measurements.
Note that the power consumption of the tag at \SI{2.4}{\giga\hertz} is still an order of magnitude lower, and at \SI{868}{\mega\hertz} two
orders of magnitude lower, than the typical
transceivers used in low-power wireless networks~\cite{CC2420}.  
%We note, the 
%TV: this would be the second not within a few lines, so better to skip it
The higher power consumption
when compared to existing state-of-the-art~\cite{HitchHike,kellogg2016passive,interscatter} is due to the use of off-shelf-components in the design
of the tag. In the future, we will implement our tag on IC to reduce the power consumption.

%\fakepar{Energy harvesting}
%\input{energy}

\begin{figure*}[t]
\centering
\includegraphics[width=\textwidth]{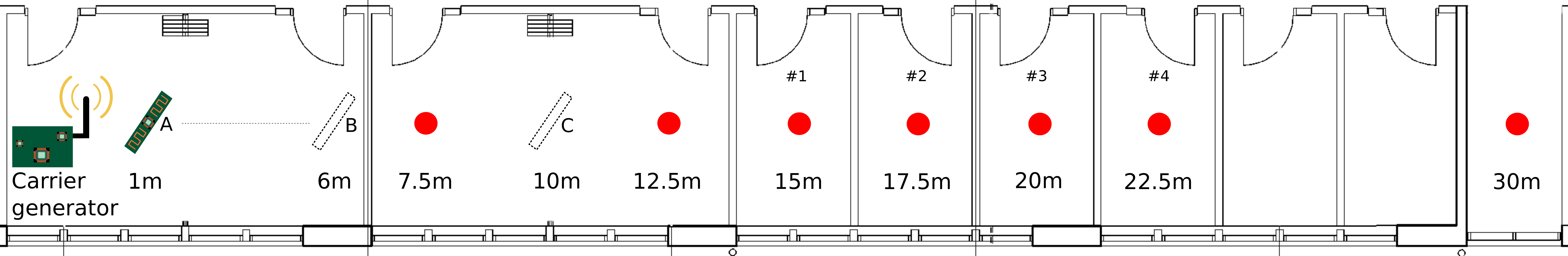}
\caption{\capt{Layout for the indoor experiments ( \SI{2.4}{\giga\hertz} ).} The carrier generator is placed in the first room, while we vary the locations of the backscatter tag (A,B,C) and the receiver (red dot).}
\label{indoorlayout}
\end{figure*}

\subsection{Supporting frequency diversity }
\label{sec:freqdiversity}

%Supporting frequency diversity
%%, or the ability to operate on different wireless channels 
%is crucial to %operations for
%most wireless protocols today. 
%%For example, BLE changes operating channel frequency every XX ms. 
The ability to operate on different frequencies 
brings numerous advantages, for example, 
mitigating the harmful effects of multi-path fading, 
reducing interference, and 
improving network capacity~\cite{bahl2004ssch}.  However, state-of-the-art backscatter systems~\cite{kellogg2016passive,interscatter,blebackscatter}
demonstrate the ability to generate transmissions on a specific
frequency. Hence, a key and unsolved 
challenge is to enable the ability
to change the frequency of backscatter transmissions.

\subsubsection{Realising frequency diversity}\hspace*{\fill}

To support frequency diversity, 
Equation~\ref{eq:mixing_equation} shows that there are two parameters that
determine the channel frequency of the backscatter transmission:
$\Delta f$, that is controlled by the tags, and $f_c$, that is
controlled by the carrier generator.
%To change
%the frequency of the backscatter transmissions, we have thus two options:
%Either the tags choose the channel by selecting an 
%appropriate value of $\Delta f$, 
%while we keep the carrier at a constant frequency; 
%or the carrier generator
%selects a frequency by adapting $f_c$.
%while the tags keep their $\Delta f$ constant. 
Changing the frequency at the tag has the following drawbacks: 

\fakepar{Tag complexity and energy consumption} Setting the operating
frequency at the tag might significantly increase the complexity of the
tag's design. 
%The only option to select the operating frequency is to increase $\Delta f$ because the tag must place the backscattered signal at some minimal offset away from the carrier to leverage the transceiver's ability to reject self-interference. 
Both, increased complexity and larger $\Delta f$ will lead to higher
energy consumption. The larger range of $\Delta f$ will
further lead to an increase in the dynamic power
dissipation.%~\cite{onbodybackscatter}.

%Changing frequency at the tag has
%the following drawbacks:  
%First, selecting the operating frequency at the tag 
%might significantly increase the complexity of the tag 
%which leads to increased energy consumption.
%and prevents the tags from being battery-free.
%To leverage the ability
%of transceivers to reject interference from the carrier signal,
%the tags must place the backscattered 
%signal at some offset away from the carrier. In our architecture this is
%\SI{2}{\mega\hertz} in the ISM band.
%However,  the tags can operate at \SI{2.4}{\giga\hertz}
%over a bandwidth of \SI{83.5}{\mega\hertz}. In order to select 
%any frequency within the designated band,
%however, the tags would need to vary their frequency over a much wider range
%of $\Delta f$. 
%For example, if the carrier signal is at \SI{2400}{\mega\hertz}, to have backscatter
%transmissions on \SI{2483}{\mega\hertz}, the backscatter tags need to generate a $\Delta f$ of \SI{83}{\mega\hertz}. 
%An increase in $\Delta f$ implies an
%increase in the dynamic power dissipation at the tag~\cite{onbodybackscatter}, along with the added
%complexity of dynamically changing the frequency over that range.

\fakepar{Out-of-band interference} Large frequency shifts can also cause
undesired interference even outside the intended ISM band.  As discussed
earlier, the backscatter tags reflect the carrier signal and shift it to
the desired frequency. They, however, also shift any other transmission
that occurs in the adjacent frequencies. As a result, any third-party
wireless transmissions will also be shifted by $\Delta f$. We
illustrate this in Figure~\ref{fig:spectrumshift} on the example of an unmodulated carrier and a WiFi transmission. The figure shows the backscatter transmissions at the desired frequency offset, but it also depicts that the WiFi transmission is shifted to the unregulated spectrum (indicated as shaded red area).
Together with the above observation, we can conclude that $\Delta f$ should be kept as small as possible, which is not compatible with changing the frequency at the tag.

%Second, even if the tags could achieve large frequency shifts at reasonable
%energy cost and complexity, this also introduces a new problem. \emph{Large 
%frequency shifts can cause unnecessary undesired  interference even outside
%the intended ISM band}.  As discussed earlier, the backscatter tags
%reflect the carrier signal and shift it to the desired frequency.
%They, however, also shift any other transmission that occurs in the adjacent
%frequencies. As a result, any third-party wireless transmissions 
%while the tag is transmitting will also be shifted away by
%an amount equal to $\Delta f$. These undesired shifted images could
%cause interference on other co-located systems, on the receiver
%itself, or they could entirely fall outside of the intended frequency
%band into the regulated spectrum. We illustrate this in
%Figure~\ref{fig:spectrumshift}. In Figure~\ref{fig:spectrumwifionly}, 
%we start a WiFi transmission. Next, we  generate an unmodulated carrier signal, 
%that we backscatter 
%at $\Delta f$ of \SI{0}{\mega\hertz}.\tv{looks like something went wrong here, probably not a 0 shift}
%The figure shows the expected
%backscatter transmissions at the desired frequency offset. It also
%depicts that the WiFi transmission is shifted to the unregulated
%spectrum (indicated as shaded red area). One way to avoid 
%the impact of this unintentional shifting of ambient traffic is to 
%keep $\Delta f$ as small as possible, 
%which is not compatible with changing the frequency at the tag. 

\fakepar{Lack of carrier sensing} One of the advantages of changing the
operating frequency is to mitigate harmful effects of cross-technology
interference which requires carrier sensing. However, passive receivers
most commonly employed on backscatter tags are not frequency 
selective, and thus lack the 
functionality to perform carrier sensing~\cite{interscatter}. As a result, backscatter tags are unable 
to decide the least interfered frequency to operate on. 

%Finally, one of the advantages of changing the operating frequency is
%to mitigate harmful effects of cross technology interference which requires
%the need to perform carrier sensing. However, 
%backscatter tags lack the functionality to perform carrier sensing,
%and thus they are not able to select the least interfered frequency. Existing backscatter
%tags even with an envelope detector do not have
%the necessary channel selectivity to discern among 
%individual channels.  As a
%result, the tags are unable to decide
%the least interfered channel to operate on. 

%\emph{We advocate changing the operating frequency of the carrier signal
%to induce change in the frequency of backscatter transmissions to
%keep complexity and energy consumption low at the backscatter tag.}. 
 
\emph{Therefore we advocate that to change the frequency of the
backscatter transmissions, we change the frequency $f_c$ of the carrier
signal rather than the frequency offset $\Delta f$ the tags induce when
backscattering.} 
This keeps the backscatter tag's complexity and energy consumption low, limits out-of-band interference, 
and allows for informed channel selection to avoid CTI~(see Section~\ref{avoidintf}). 

\subsubsection{Unison backscatter}\hspace*{\fill}

Almost anywhere we are, we are surrounded by several commodity
devices. For example, we might have sensor nodes or WiFi access 
points as part of the infrastructure or we carry
fitness trackers that are equipped with BLE radios. 
Interscatter~\cite{interscatter} demonstrated that
BLE radios can generate a carrier signal and that this carrier can be backscattered
as a WiFi signal at a fixed frequency.   However, backscatter signals
are inherently weak and are prone to interference from ambient
wireless traffic. We next present a technique we call \emph{Unison backscatter} which
helps improve reliability when operating in interfered environments.

We build Unison by borrowing concepts from MIMO; receiving with multiple receivers on
separate frequencies helps to improve reliability.
We use several devices to generate carrier signals at different frequencies.
Because of the mixing property at the backscatter tag this leads to
simultaneous transmissions at all the frequencies. 
%If we enable several 
%commodity devices, with each programmed to 
%generate carrier signal
%at separate frequency,  
%the mixing property at the backscatter tag would result in having
%simultaneous  transmissions at all these frequencies.
% --- last minute remove of equation, integrate it in the text. 
%For example, if we have carrier signals
%at frequencies $f_{c1}$, $f_{c2}$ and $f_{c3}$, we get backscatter transmissions
%on the following frequencies (assuming we discard the mirror images $m(t)$
%from the mixing operation):\chroh{I am not in favour of a equation omitting a part. Added $+m(t)$ for the mirror images.} 
%\begin{equation}
%\begin{split}
%(\sin(f_{c1} t) +   \sin(f_{c2} t) + \sin(f_{c3} t)) \sin(\Delta f t) \\ = \cos[(f_{c1} + \Delta f)t]  + \cos[(f_{c2} + \Delta f)t]  + \cos[(f_{c3} + \Delta f)t] + m(t)
%	  \label{eq:unison_equation}
%\end{split}
%\end{equation}
%The equation above shows that we can generate shifted backscatter
%transmissions at a $\Delta f$  offset from each of the 
%three carrier frequencies. By having multiple receivers at the reader we
%can improve its reliability since it is sufficient if any of the three receivers receives the backscattered data.
For example, if we have carrier signals
at frequencies $f_{c1}$, $f_{c2}$ and $f_{c3}$, we get backscatter transmissions
at $f_{c1}+\Delta f$, $f_{c2}+\Delta f$ and $f_{c3}+\Delta f$, respectively (assuming we discard the mirror images
from the mixing operation).
By having multiple receivers at the reader we
can improve its reliability since it is sufficient if any of the three receivers receives the backscattered data.

%by ensuring that if any of 
%the transmission is lost due to to interference, we can recover information from transmissions r%eceived by the other receivers.
While we demonstrate \emph{Unison backscatter}  for our architecture, the technique is equally applicable to
other backscatter systems.  

In using multiple devices to generate carrier signals, or to receive transmissions,  Unison backscatter is similar to a technique 
presented by Zhang et al.~\cite{zhang_enabling_2016}. They use multiple commodity devices to improve the SNR of the backscattered
signal, while we enable concurrent transmissions on multiple frequencies at the same time. Generating carriers with multiple devices inherently increases the energy consumption for carrier generation.
The devices generating the carrier are, however, usually 
more powerful and might also be powered externally.

\subsection{Cost analysis}
\label{costanalysis}
We implement our architecture using off-the-shelf components.  We next present
the overall cost of our architecture\footnote{We designed
a few lab prototypes for the experiments conducted in this paper. 
We expect the overall cost to be substantially lower when produced 
at scale.}.

\fakepar{Backscatter tag} The  tag 
is designed using Autodesk Eagle software 
and ordered at OSH Park at a cost of 5 USD
for three boards. The RF switch
costs 2.5 USD, one ultra-low power oscillator
costs 1.8 USD~(3.6 USD for two), and the multiplexer costs 2.6 USD resulting in an overall cost of the
tag around 10.3 USD.

\fakepar{Reader} We implement the \SI{2.4}{\giga\hertz} reader
using a CC2500 transceiver module from MikroElektronika interfaced to an
Arduino Zero platform. The radio module costs around 20 USD
and the Arduino Zero approximately 
50 USD. The overall cost
of the \SI{2.4}{\giga\hertz}
reader is hence approximately 70 USD.
We implement the \SI{868}{\mega\hertz} reader using a Texas
Instruments CC1310 launchpad board~\cite{cc1300launchpad} that costs around 29 USD.

\fakepar{Carrier generator} A key feature of our architecture is its ability
to use wireless devices that are part of the existing infrastructure to generate the carrier signal
incurring no additional cost. If needed we can also use the Texas Instruments
CC3200 launchpad board~(\SI{2.4}{\giga\hertz})~\cite{cc3200launchpad} or CC1310 Launchpad board~(\SI{868}{\mega\hertz})~\cite{cc1300launchpad}
that cost around 29 USD  to generate the carrier signal as we demonstrate in Section~\ref{tonegeneration}.

\section{Evaluation}
\label{evaluation}

\label{sec:eval}

In this section, we present experimental results
to evaluate different aspects of our architecture. We perform the experiment
in a range of environments and conditions. In our experiments, we  find:

\begin{itemize*}

\item In an indoor environment,  with the tag co-located with the carrier source, 
      we can communicate tens of meters
	  even when  the tag and the reader are separated by walls. 
	  When operating at \SI{868}{\mega\hertz}, we 
	  can communicate through multiple floors. 

\item In an outdoor environment,  we can communicate over distances longer than \SI{3.4}{\kilo\meter} at
   	  \SI{868}{\mega\hertz}, and \SI{225}{\meter} at \SI{2.4}{\giga\hertz} with colocated tag and carrier
   	  source, which is an order
   	  of magnitude longer than state-of-the-art backscatter systems.

\item We can leverage multiple WiFi and 802.15.4 radios to provide the carrier signals at distinct frequencies to enable operations even in busy wireless environments 
	  by enabling concurrent transmissions on multiple wireless channels.

%\item As opposed to conventional radios, who change frequency. In backscatter device, we demonstrate changing frequency at the carrier generator is apt to reduce interference.

\item We demonstrate that changing the frequency at the carrier generator (rather than changing the frequency offset at the backscatter tag) provides frequency diversity which increases reliability under external interference.

\end{itemize*}

\subsection{Range and Bit Error Rate}
\label{rangeandber}
% vim: textwidth=72 spell spelllang=en_us

We first aim to understand the achievable range and reliability of
our architecture in different environments and operating modes.

\fakepar{Experimental setup}  We equip both the carrier generator
and the tag with omnidirectional antennas. For experiments
at \SI{2.4}{\giga\hertz} we employ TP-Link~\cite{tplink} antennas,
and at \SI{868}{\mega\hertz} we use VERT900~\cite{vert900}
antennas. At the receiver, we use an onboard inverted-F
antenna.  We mitigate the non-uniform radiation pattern  of the receiver onboard antenna
by orienting the antenna towards the tag which improves the signal-to-noise ratio~(SNR) of
the received signal. To account for different antenna orientations 
and multi-path fading, we perform three independent runs of 
each experiment.

We generate a carrier signal with a strength of apporximately \SI{26}{dBm} at 
\SI{2.4}{\giga\hertz} using a USRP B200 software defined radio~\cite{usrp} equipped with an
external amplifier. At  \SI{868}{\mega\hertz}, we generate a carrier of strength approximately \SI{28}{dBm} using a 
CC1310~\cite{CC1310} coupled together with an amplifier.   We note that the carrier
signal is a few \SI{}{\deci\bel}s lower than the maximum permissible under
FCC regulation, and used by other systems~\cite{kellogg2016passive, HitchHike}.
With a stronger carrier, we expect to  improve the range.
Unless otherwise stated, we position the  tag, receiver
and carrier generator 
\SI{1}{\meter} above the ground.

\fakepar{Metrics and communication parameters} In each experiment run, we
transmit 100 randomly generated packets of 64 byte and 36 byte
each for experiments conducted at \SI{2.4}{\giga\hertz} and \SI{868}{\mega\hertz} 
respectively.  
On the receiver, we keep
track of the received packet sequence number, signal strength and the noise
floor.  We collect approximately $10^5$ bits, and compare the received bits
with the transmitted bits as done in recent backscatter
works~\cite{BackFi,zhang_enabling_2016}.  We calculate the bit error rate (BER) for
each run of the experiment, along with its mean and standard deviation between
runs.  Unless otherwise stated, the backscatter tags transmit at a
rate of \SI{2.9}{kbps}.

\begin{figure}[!tb]
  \centering
  \includegraphics[width=\linewidth]{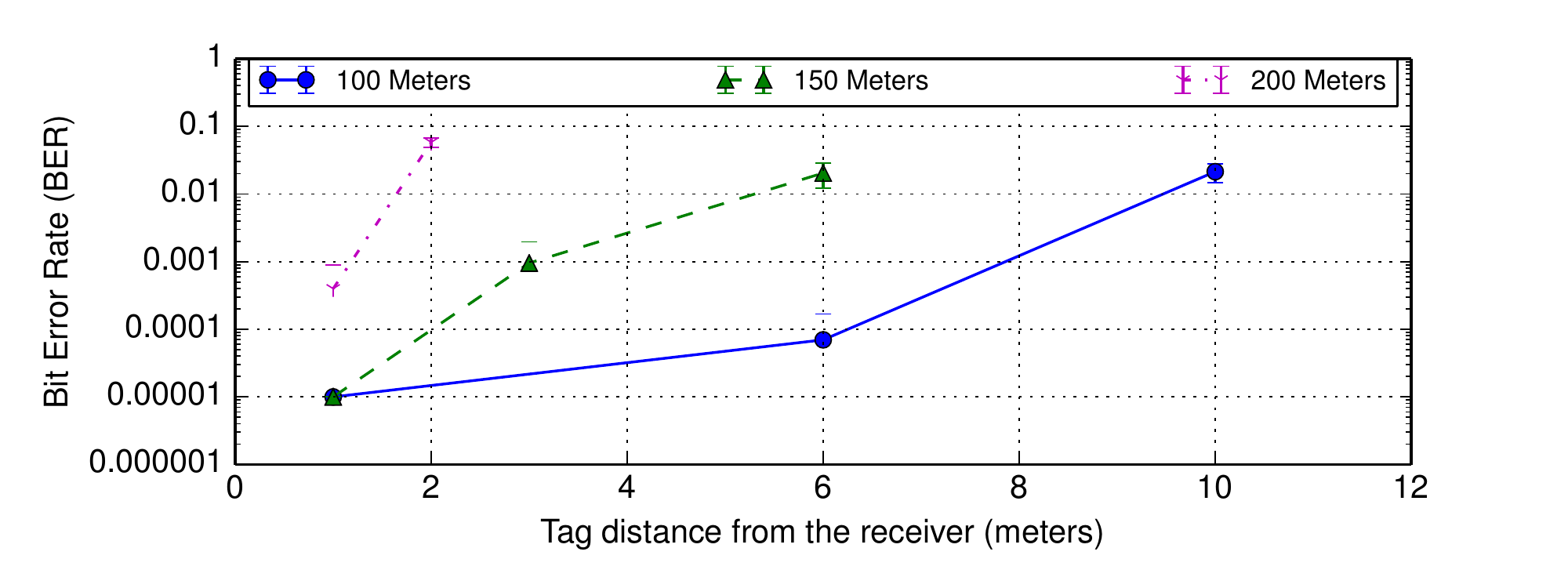}
  \caption{\capt{Backscatter tag close to the receiver (outdoors, \SI{2.4}{\giga\hertz})}. As the distance from the carrier generator increases, the maximum possible range between the backscatter tag and the reader decreases. }
  \label{backscatterclosereceiver}
  \end{figure}

\subsubsection{\SI{2.4}{\giga\hertz} architecture}\hspace*{\fill} 

\fakepar{Outdoors} We begin our evaluation outdoors with line-of-sight
propagation. The experiments are conducted outside of our university,
with buildings on one side and forest on the other side.

\begin{figure}[!tb]
\centering
\includegraphics[width=\linewidth]{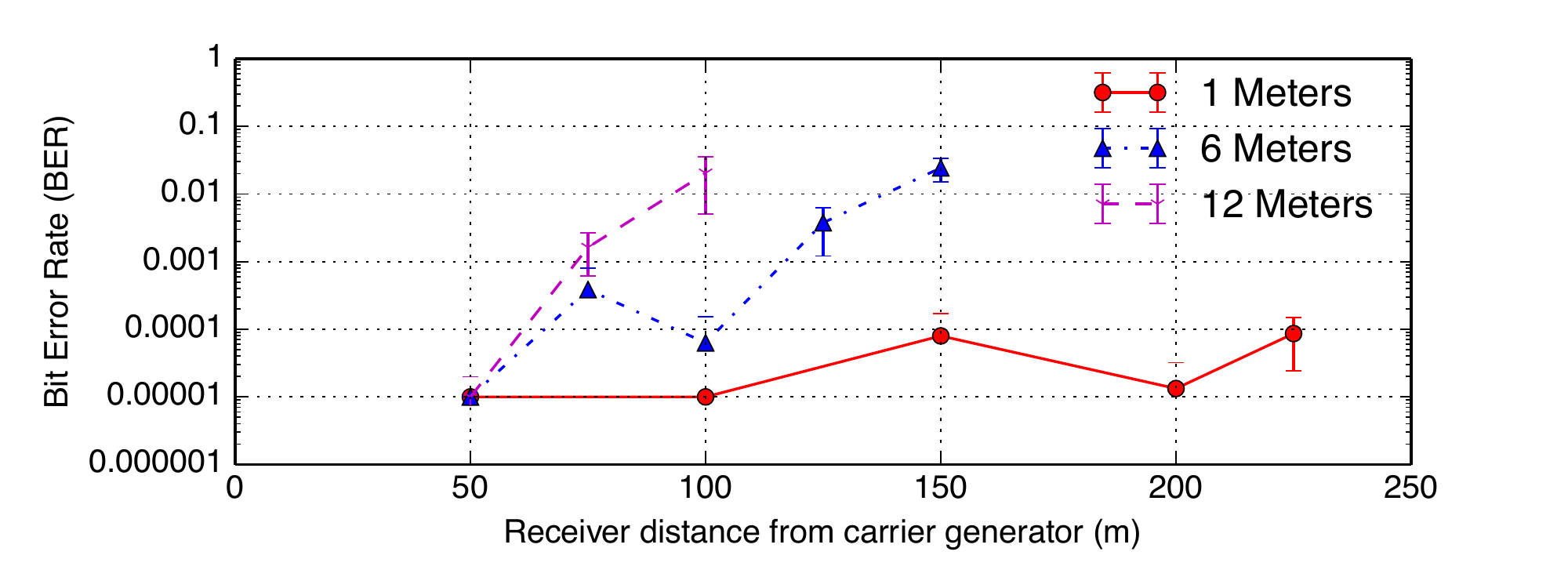}
\caption{\capt{Backscatter tag close to the carrier generator (outdoors, \SI{2.4}{\giga\hertz})}.
  Positioning the backscatter tag close to the carrier generator leads to a high range. A distance of \SI{12}{\metre}
  is sufficient to achieve \SI{100}{\metre} range at a BER of $10^{-2}$.}
\label{outdoorlowbitrate}
\end{figure}

We first assess the impact of
positioning the backscatter tag  close to the carrier generator.
Figure~\ref{outdoorlowbitrate} shows the observed BER as a function of
distance between the receiver and the carrier generator using the
CC2500-based receiver that operates in the \SI{2.4}{\giga\hertz} band.
We achieve a range of
\SI{225}{\metre}, \SI{140}{\metre}, and \SI{90}{\metre} with 
a separation of \SI{1}{\metre}, \SI{6}{\metre}, and \SI{12}{\metre} from the
carrier generator, respectively. In most cases, the BER is well below
$10^{-2}$ which is comparable to state-of-the-art backscatter systems~\cite{HitchHike,turbochargebackscatter,ambientbackscatter}. 
As the tag moves away from the carrier
generator, the achievable range decreases while the  bit errors
increase.

We next evaluate the impact of positioning the tag close to
the reader. Figure~\ref{backscatterclosereceiver} shows the result of
the experiment.  As both the tag and the reader move farther
away from the carrier generator, the communication range decreases.
When the tag is at a distance of \SI{200}{\metre} 
from the carrier generator, the reader can only
receive reliably up to \SI{2}{\metre} from the tag. However, when the distance between
carrier generator and tag is \SI{100}{\metre},
we can receive reliably even when the distance between tag and receiver is \SI{10}{\metre}.

The results of the experiment suggest that the optimal position to
achieve low BER and high range is to either position the backscatter tag
close to the carrier generator, or to take the reader close to the
backscatter tag, especially when operating at longer distances from the
carrier generator. These results correspond to the theoretical findings in 
Section~\ref{subsec:carrier_generation}. 

\begin{figure}[!tb]
\centering
\includegraphics[width=\linewidth]{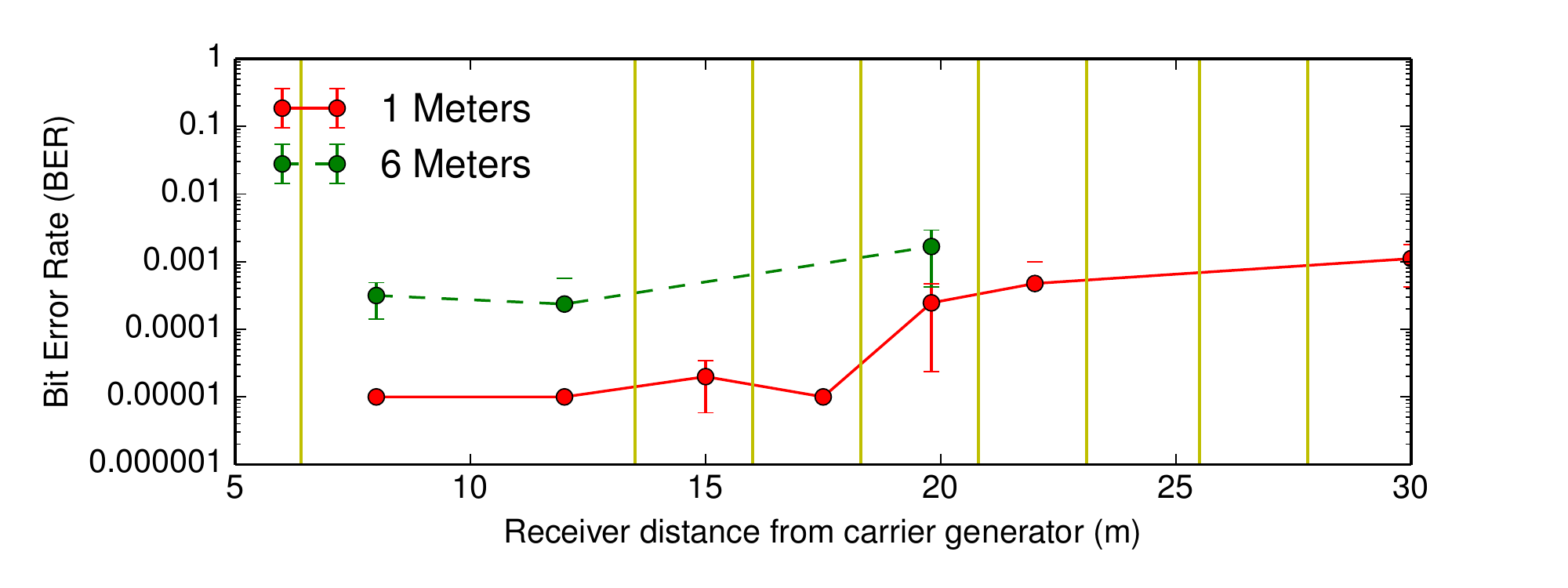}
\caption{\capt{Through the wall} (\SI{2.4}{\giga\hertz}).  The vertical lines indicate the presence of walls.
When the tag is kept 1m from the carrier generator, we can receive transmissions
eight walls away  at a distance \SI{30}{m} from the carrier generator. }
\label{throughwall}
\end{figure}

\begin{figure}[!tb]
\centering
\includegraphics[width=\linewidth]{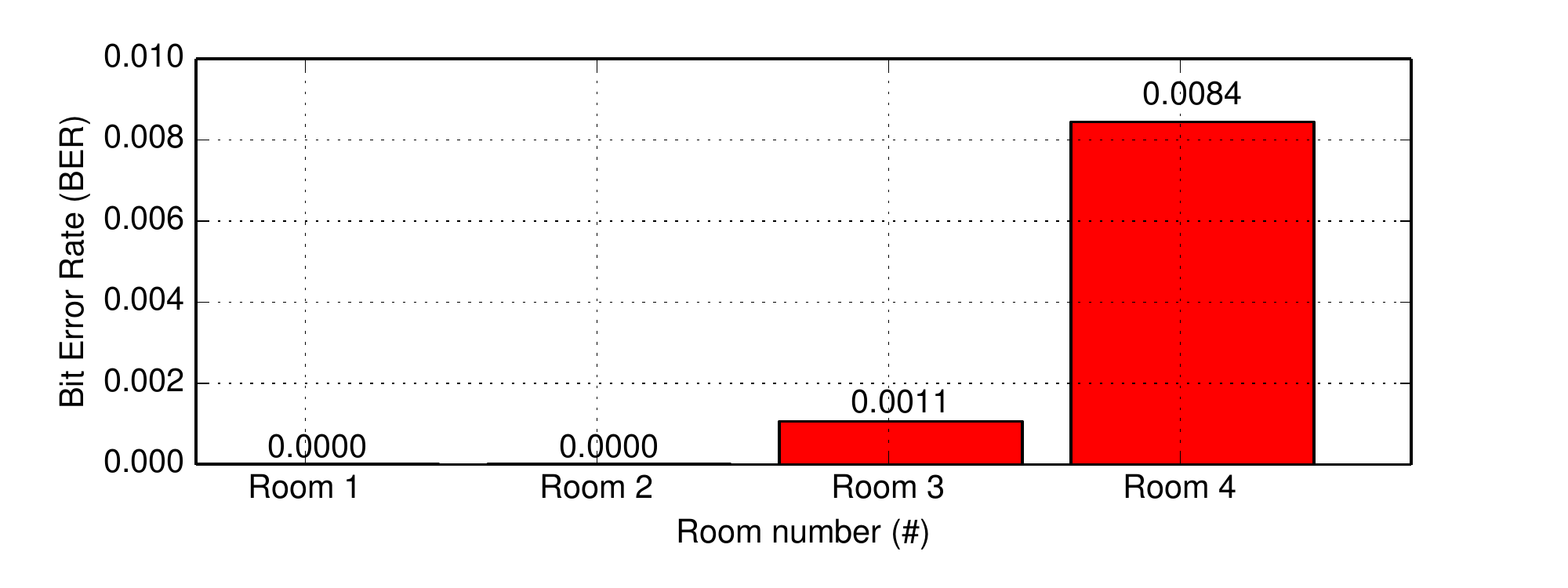}
\caption{\capt{Room to room backscatter} (\SI{2.4}{\giga\hertz}). Carrier generator and backscatter
  tag are placed in separate rooms and are separated by \SI{10}{m} (tag position C).
  We can receive transmissions even four rooms away from the tag.}
\label{roomtoroom}
\end{figure}

\fakepar{Indoors} Next, we evaluate the ability of \system to operate in
non-line-of-sight environments. We perform experiments in an indoor
environment in the presence of rich fading and other wireless networks.
The environment is shown in Figure~\ref{indoorlayout}. The study rooms are of
varying size between \SI{2.5}{m} and \SI{7.5}{m}, and each room is separated 
by an insulated gypsum wall of approximately 16~cm. The rooms are equipped 
with tables, chairs, and a whiteboard on the wall separating the rooms. 

In a first experiment, we place the backscatter tag and the carrier
generator in the same room (see Figure~\ref{indoorlayout}). 
We position the backscatter tag \SI{1}{m} and \SI{6}{m} away from the carrier generator.  
We vary the position of the receiver by placing it in different rooms.

Figure~\ref{throughwall} shows the results, where vertical lines
indicate the presence of walls in the figure. When the tag
is located at a distance of \SI{1}{m} from the carrier generator, we can achieve
a distance of approximately \SI{30}{m} between the receiver and carrier generator,
traversing through eight walls. 
At longer distances the SNR falls below the sensitivity level
of the radio.  As the distance between the tag and the carrier
generator increases to \SI{6}{m}, the strength of the backscatter signal
reduces, which affects the achieved range and also introduces higher bit
errors.  We achieve a range of approximately \SI{20}{m} with five walls
separating the tag and the receiver. 

\fakepar{Room to Room Backscatter} We next evaluate \system in a
scenario where tag, carrier generator and the receiver are all
located in separate rooms.  We keep the carrier generator in the same
location as in the earlier experiment, and move the tag to the next
room (tag position C).  The distance between the tag and the carrier
generator is \SI{10}{m}, and a wall separates them. We place the receiver in
different rooms and repeat the experiment.

Figure~\ref{roomtoroom} shows the result of the experiment.  We can
receive backscattered transmissions four rooms away from the backscatter
tag with four walls separating the backscatter tag and the receiver at a BER
lower than $10^{-2}$. 
We note that existing CRFID systems do not operate well in through-the-wall
scenarios~\cite{turbochargebackscatter}. Hence, we believe that \system's
ability to perform well in through-the-wall scenarios  is a significant
improvement.

\begin{figure}[!tb]
\centering
\includegraphics[width=\linewidth]{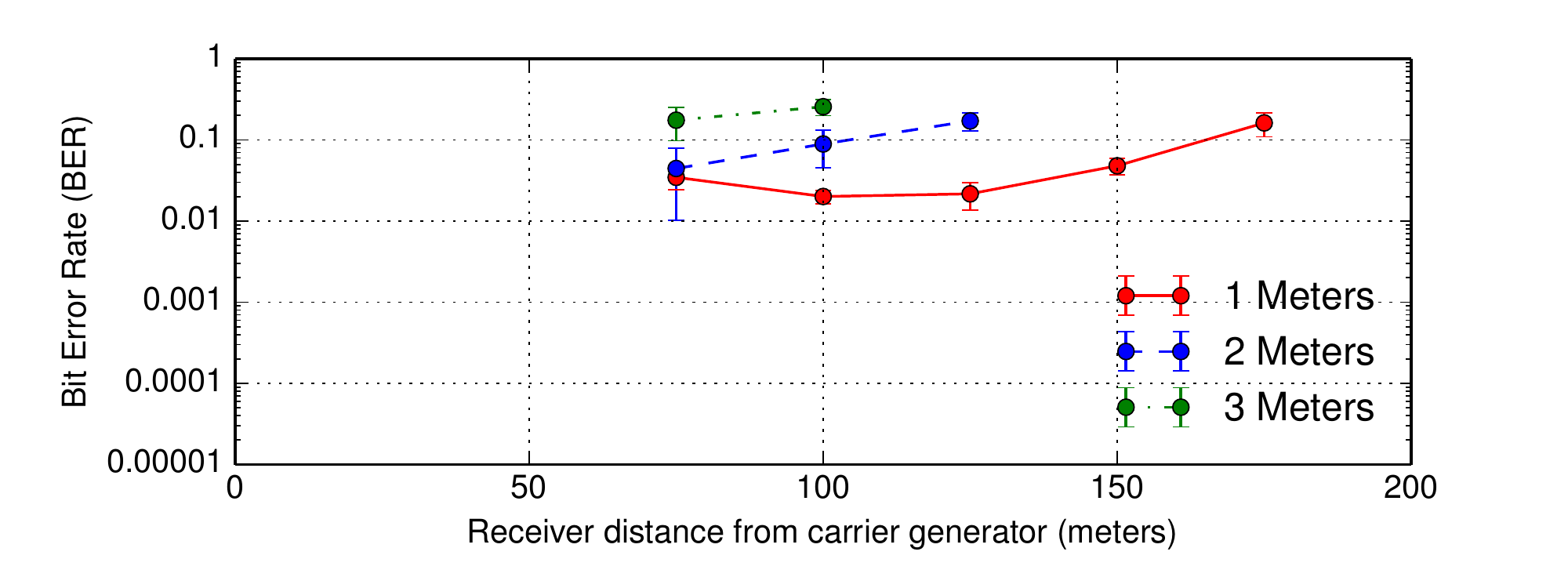}
	\caption{\capt{High bitrate (197 $kbps$) (\SI{2.4}{\giga\hertz}, outdoors).} A High bitrate
												reduces the achievable range and introduces higher
												bit errors as opposed to operating the reader at
												lower bitrates. The tag was co-located with the carrier generator.} 
											\label{outdoorhighber}
\end{figure}

\fakepar{High-speed mode}  Some sensing applications such as battery-free cameras~\cite{wispcam} or
microphones~\cite{thomas2013rich}, suffer from the low bitrates of
CRFID. To support such applications, \system supports higher bitrates at
the cost of reduced receiver sensitivity. We next perform an experiment outdoors
to investigate this trade-off. We program the reader and the receiver
to operate at a bitrate of 197 $kbps$ at \SI{2.4}{\giga\hertz},
which is close to the maximum
achievable goodput of IEEE 802.15.4~\cite{Varshney:2015:DTR:2809695.2809720}, a widely used protocol in wireless sensor networks.
\label{sec:highspeed}

We position the  tag close to the carrier generator at
distances of \SI{1}{}, \SI{2}{} and \SI{3}{\metre}, and place the reader at intervals of
\SI{25}{\metre} starting at a distance of \SI{75}{\metre} from the carrier generator.
Figure~\ref{outdoorhighber} shows the result of the experiment. While we
achieve a range of \SI{100}{\metre} at a target BER of $10^{-2}$ when the
 tag is located \SI{1}{\metre} apart from carrier generator, the BER
increases significantly at larger distances. 

The observed BER is %orders in magnitude higher 
significantly higher
than at low bitrates at
similar distances.  However, the BER 
we achieve is comparable
to the recent backscatter systems operating 
at similar bitrates and frequency, while we get a nearly 
threefold improvement in range~\cite{HitchHike}. 
The experiment suggests that high-speed mode should
only be used at short distances or together with suitable mechanisms at
the reader to recover lost or corrupt bits, or to improve the reliability
of links using error correction and bit spreading 
mechanisms as we describe in our
recent work~\cite{ambujhotwireless}.

%\begin{figure}
%    \centering
%    \includegraphics[width=\columnwidth]{results/subghz.pdf}
%    \vspace{-6mm}

%    \caption{\capt{Backscatter tag close to carrier generator (outdoors, \SI{868}{\mega\hertz})}.
%      We can achieve a range of \SI{475}{\meter} with a low BER with tag separated from carrier generator by \SI{1}{\meter}.}
 %     \vspace{-4mm}

%    \label{fig:outdoorsubghz}
%\end{figure}

\begin{figure}[!tb]
\centering
\includegraphics[width=0.7\linewidth]{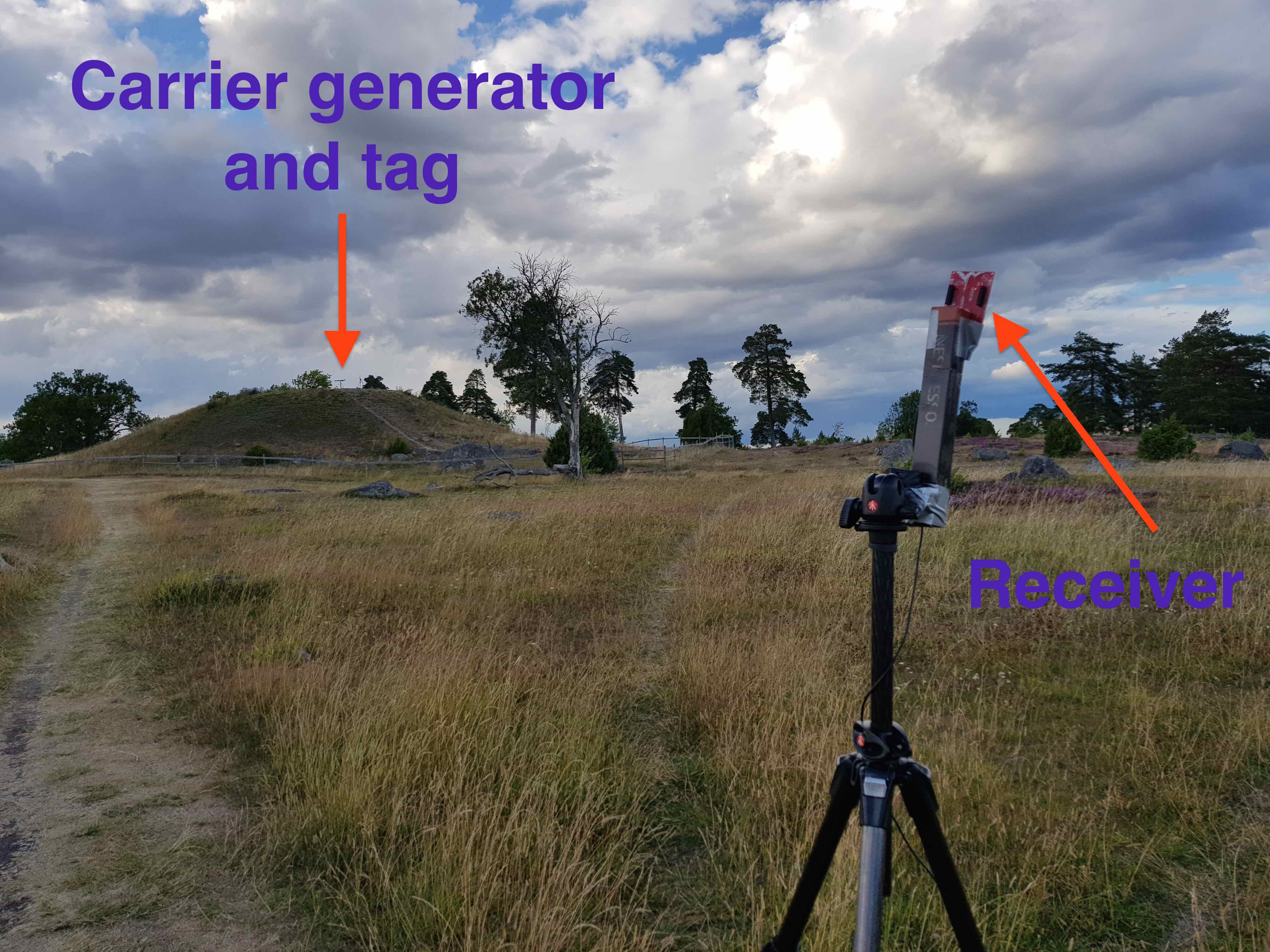}
\caption{\capt{Long distance backscatter.} Carrier generator and tag were co-located on a small plateu a few meters above the ground. The receiver
     was approximately \SI{2}{\meter} above the ground on a tripod.}
\label{longdistance}
\end{figure}

\begin{figure}[!tb]
\centering
\includegraphics[width=\linewidth]{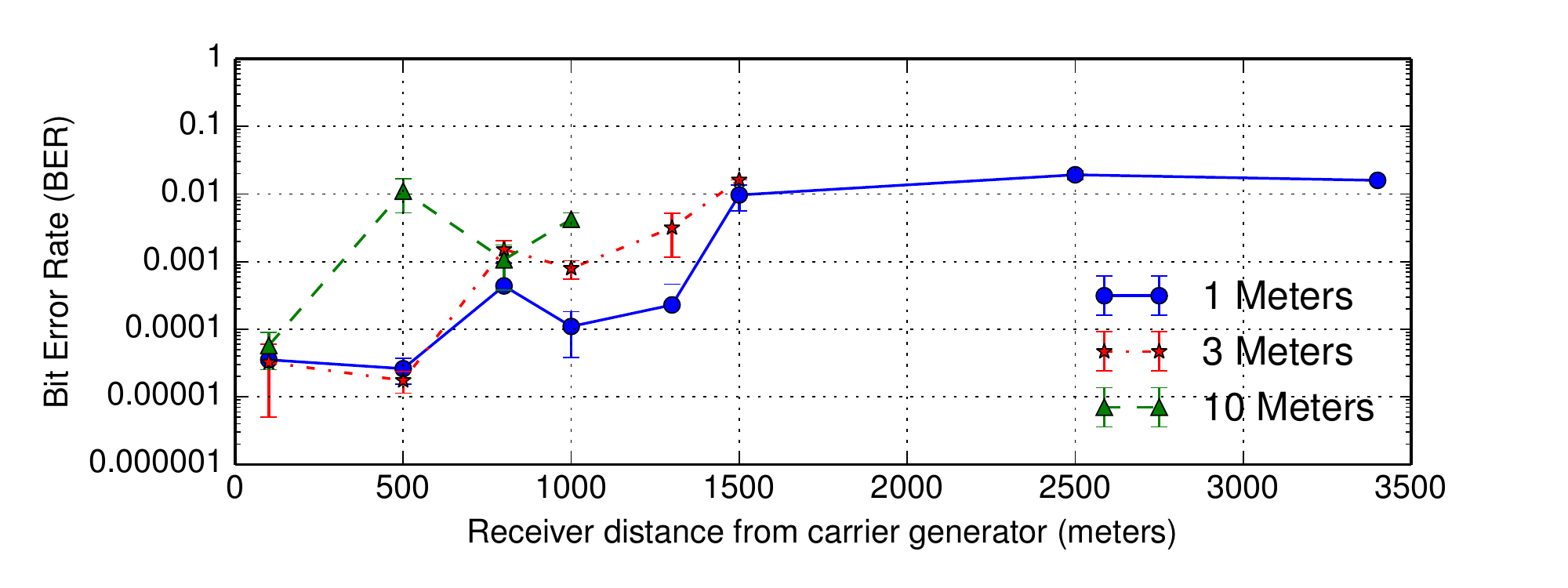}
\caption{\capt{Long range backscatter (outdoors, \SI{868}{\mega\hertz})}. Even when the carrier generator
and tag are located \SI{10}{\meter} apart, we can communicate to distances as high as \SI{1}{\kilo\meter}. At a distance
of \SI{1}{\meter} between tag and carrier source, we can communicate to distance of \SI{3.4}{\kilo\meter}}
\label{riverexp}
\end{figure}

\subsubsection{\SI{868}{\mega\hertz} architecture}\hspace*{\fill} 

\system when operating at \SI{868}{\mega\hertz} enables
higher range and  preserves 
compatibility with existing CRFID platforms like WISP. 
While the focus of our work is to use \SI{2.4}{\giga\hertz} which
enables the use of commodity wireless devices as carrier
generators, we present 
encouraging results the architecture achieves
at \SI{868}{\mega\hertz}. For brevity, we present results where the tag close is to the carrier generator,
and the maximum range achieved
with the tag equidistant between the carrier generator
and the receiver.

\fakepar{Co-located tag and carrier generator (Outdoors)}  In this experiment, we investigate 
the maximum  range achievable with our architecture in an outdoor line-of-sight environment.  We perform an experiment similar to
the one performed earlier at \SI{2.4}{\giga\hertz}.   We perform the experiment in a large open space with some trees and vegetation. We co-locate the carrier generator
with the backscatter tag \SI{1}{\meter} above ground on a small plateau of a few meters height~(as shown in Figure~\ref{longdistance}). We keep the receiver
on a tripod approximately \SI{2}{\meter} above the ground. We position the
backscatter tag at a distance of \SI{1}{}, \SI{3}{} and
\SI{10}{\meter} from the carrier generator.

Figure~\ref{riverexp} demonstrates
the result of the experiment. At a distance of \SI{1}{\meter} between carrier source and
tag, we can receive transmissions \SI{3.4}{\kilo\meter} away. At this distance the received signal strength is close to the sensitivity level of the receiver and requires orientation of the antenna to maximize the SNR. The bit error rate is still moderate around 1.5\%. 
%At distances upto \SI{2.8}{\kilo\meter}, we could receive transmissions without careful positioning of the receiver antenna. However, at higher distances we carefully orient the antenna towards the tag to maximize the SNR and receive weak backscatter transmissions. \chroh{added a sentence above instead}
Similarly, we can communicate upto a maximum 
distance of \SI{1.5}{\kilo\meter} and \SI{1}{\kilo\meter} when the distance between the tag and carrier generator is \SI{3}{\meter} and \SI{10}{\meter} respectively. 
We observe slightly anomalous results at a distance of \SI{800}{\meter} due to the presence
of a large tree. To the best of our knowledge, this is the highest range demonstrated with backscatter communication and significantly advances the state of the art.

\fakepar{Co-located tag and carrier generator (indoors)} We also evaluated our system in the indoor environment. We place the 
carrier generator and the tag separated by \SI{1}{\meter}. As we get substantially longer
range than when operating at \SI{2.4}{\giga\hertz}, we perform the experiment at the basement
of our university, and on different floors directly above the tag.
In the basement, in most locations the tag and the receiver were not in line-of-sight.

\begin{figure}
    \centering
    \includegraphics[width=\columnwidth]{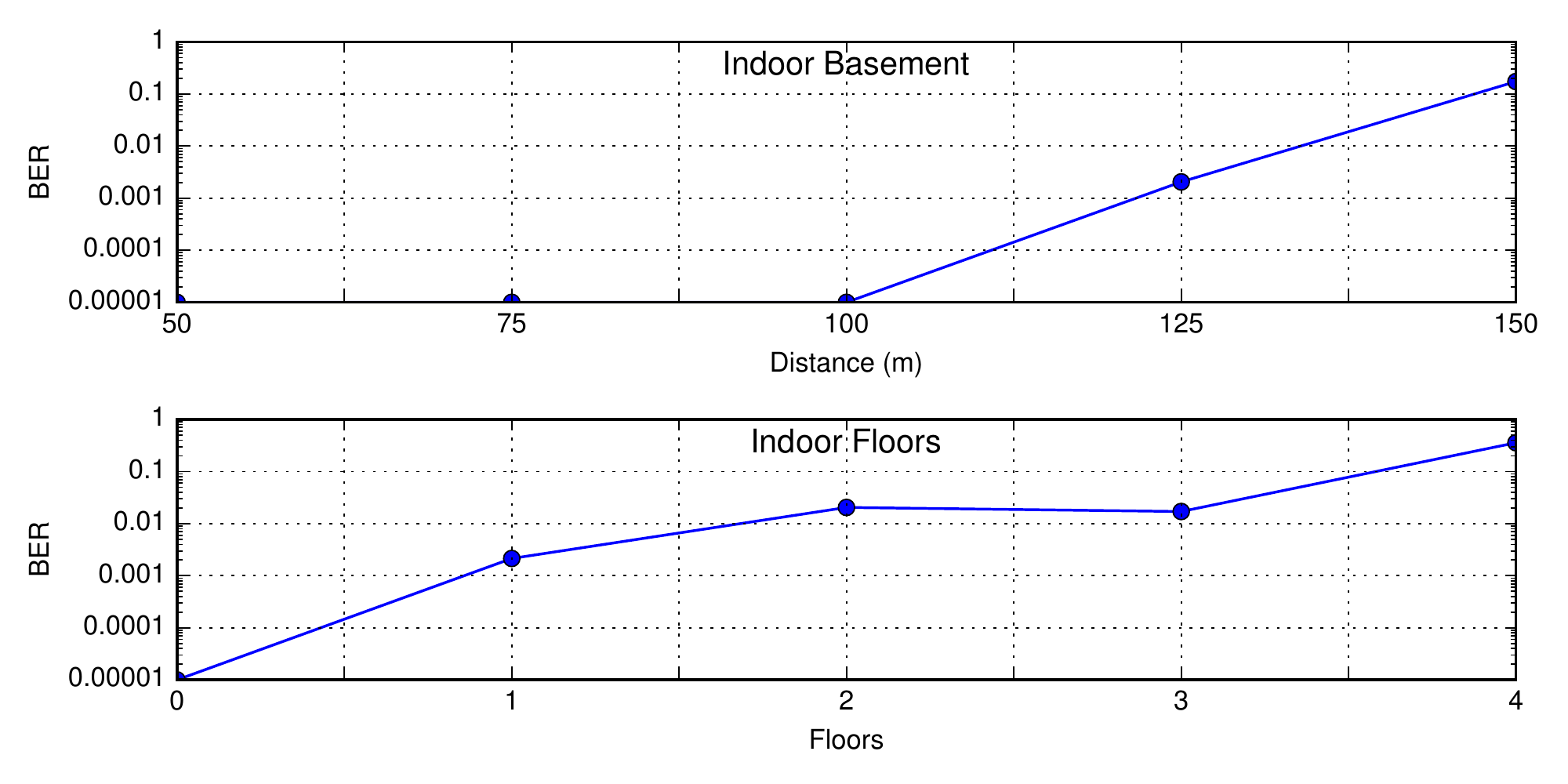}
    \caption{\capt{Indoors (\SI{868}{\mega\hertz})}.
      We can communicate through several floors of the university building. The tag and carrier generator are separated by \SI{1}{\meter}.}
    \label{fig:indoorsubghz_floors}
\end{figure}

Figure~\ref{fig:indoorsubghz_floors}  shows the results of experiment. The figure shows that we can communicate over
multiple floors of the building (up to the 4th floor). The BER increases sharply with the number of floors, 
as the SNR of the signal becomes progressively worse. We note, other backscatter systems~\cite{HitchHike, fmbackscatter}
also exhibit such sharp increase in BER when the distance increases. In the basement, we can reach a distance of \SI{150}{meters}. To the best of our knowledge, no existing backscatter
system has been able to demonstrate the ability to communicate through multiple floors in the building.

\fakepar{Tag equidistant between carrier generator and receiver}  Finally, we perform the experiment with the backscatter tag equidistant between the carrier generator and the receiver.  
As discussed in Section~\ref{subsec:carrier_generation}, 
this configuration results in the weakest received signal strength and hence communication range.

%We perform the experiment by positioning the backscatter tag equidistant from both the carrier generator and the receiver. \chroh{just mentioned two sentences ago}
We position the tag in line-of-sight with both the carrier generator and the receiver and find the maximum separation that 
achieves signal levels close to the transceiver's sensitivity level.
In our experiment, we can keep the tag
a maximum distance of approximately 
\SI{75}{\meter} from both the carrier generator and the receiver. Our experiment suggests that our architecture when operating in monostatic mode can achieve
a communication range as high as \SI{75}{\meter}. This is because
the particular configuration has similar path loss to the monostatic 
configuration of RFID readers, and hence represents a 
significant improvement over RFID readers that communicate only up to a
maximum distance of \SI{18}{\meter}~(See Section~\ref{crfid}).

%We note, as such a configuration
%represents the weakest received signal levels, the results suggests 
%%that the tag could be placed anywhere between the carrier generator
%and the receiver 
%\SI{150}{\meter}, the tag could be placed anywhere in between
%and yet be able to communicate.

\subsection{Leveraging Carriers from Existing Infrastructure}
\label{tonegeneration}
% vim: textwidth=72 spell spelllang=en_us

%\begin{figure}[!tb]
 %   \centering
  %  \includegraphics[width=.6\columnwidth]{figures/multitag-setup.pdf}
  %  \vspace{-2mm}
  %  \caption{\capt{Spatial setup for distributed-carrier setup.}}
  %  \label{fig:eval:dist-carrier:setup}
  % \vspace{-4mm}

%\end{figure}

\fakepar{Simultaneous carrier from commodity radios} In this experiment, we investigate the
impact of generating a carrier from multiple devices at the same frequency. We
deploy six MSP430-based backscatter tags in an office in close
proximity to TelosB sensor nodes~\cite{polastre2005telos}.  Their radio chips 
(CC2420) feature a test mode that
allows to generate an unmodulated carrier at an output power of \SI{0}{dBm}. 
The tags
periodically backscatter packets with random payloads.
We place a CC2500-based receiver in the same  room.  We collect 
received packets over a time span of five hours.

\begin{figure}[!tb]
    \centering
    \includegraphics[width=0.8\columnwidth]{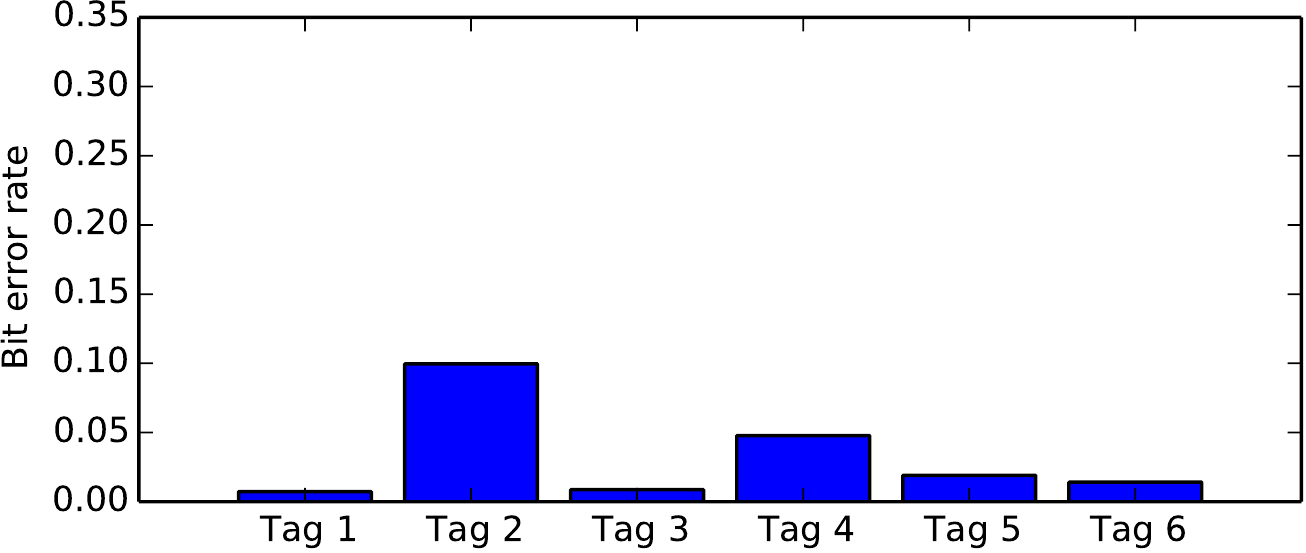}
    \caption{\capt{Bit error rate for distributed-carrier
        setup.} \system can make use of several carriers from a deployed infrastructure.}
    \label{fig:eval:dist-carrier:ber}
\end{figure}

Fig.~\ref{fig:eval:dist-carrier:ber} shows the BER for
each of the six tags. BERs are generally low, except for tag
2, which has the longest distance to the receiver. We attribute the bit
errors that we observe to interference from other coexisting wireless
networks and occasional collisions between
transmissions from tags. We do not observe distinct temporal variations
in the bit error rate. We conclude that using several carriers
simultaneously is feasible, and that slight offsets in the carrier
frequency between carrier generators (which are inevitable due to
variations in crystals) do not noticeably affect communication.

\fakepar{Range with commodity radios} In the next experiments, we investigate
the range we can achieve when using commodity radios to generate the
carrier signal. As the transmit power of these radios is much lower than those of
SDRs, we expect the range to decrease.  We keep the tag \SI{1}{\metre} 
from the carrier generator. We perform the experiments outdoors, in line-of-sight.

First, we perform an experiment with the CC3200 WiFi transceiver.
We use the CC3200 Launchpad~\cite{cc3200launchpad} (\$30)
as platform. The CC3200 transmits at its maximum
power of \SI{18}{dBm} .  We achieve a range of \SI{54}{\metre}. We expect wireless devices such as WiFi routers that use this particular transceiver, or operate at similar
output power  when used as carrier generator in our architecture 
to result in similar range.
Next, we perform an experiment with the TelosB sensor node (\$70)
using its 802.15.4 radio as carrier generator.  The sensor node
transmits at \SI{0}{dBm} power. A lower carrier strength
results in a maximum range of  \SI{7.5}{\metre}.

On both platforms, we only use 
the on-board antennas that have a limited gain  
which limits the achievable range. Despite the much lower carrier
strength, our architecture is able to achieve
a range that is comparable with the state of the art (see Table~\ref{table:comp}).

\subsection{Unison backscatter}

\begin{figure}
    \centering
    \includegraphics[width=\columnwidth]{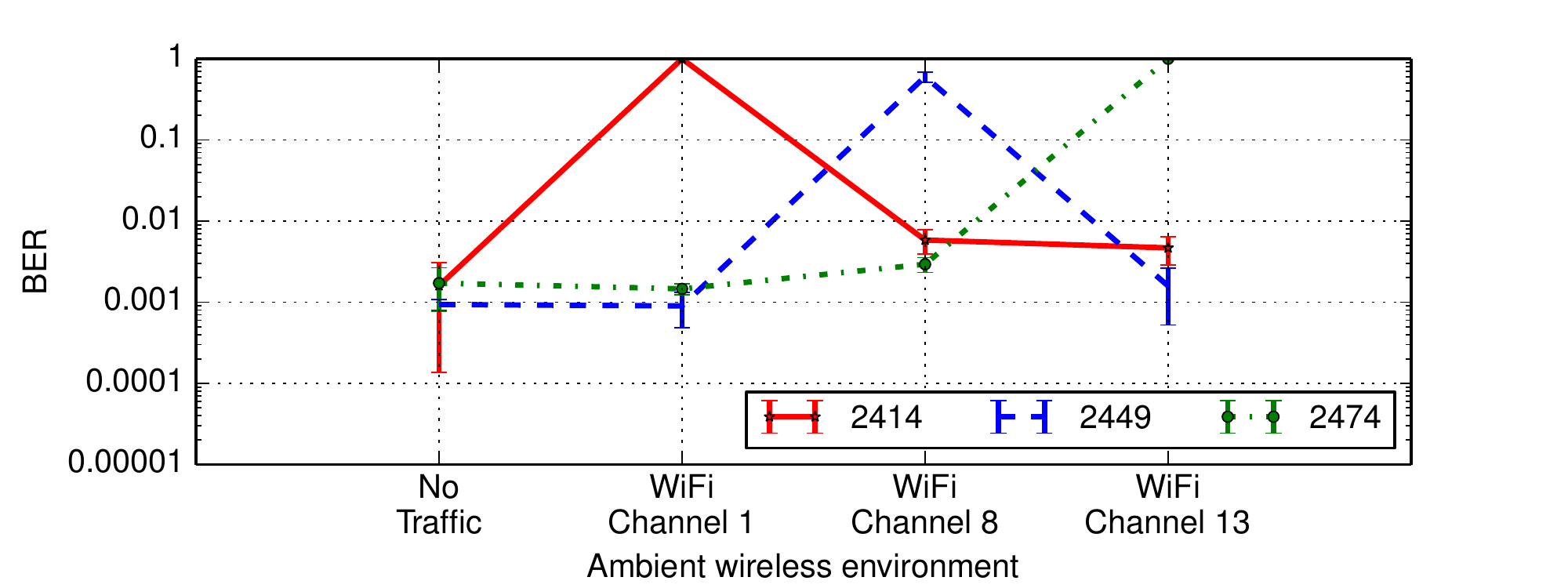}
    \caption{\capt{Unison backscatter}. Leveraging multiple commodity devices to generate the carrier signal, together with multiple receivers at the reader 
    keeps bit errors low even under external interference.}
    \label{unisoneval}
\end{figure}

%In this experiment, we investigate if several commodity devices
%together generating carrier signal at distinct frequencies, coupled together with multiple
%receivers could help to improve reliability when receiving
%backscatter transmissions under interference.
In this experiment, we investigate the \emph{Unison} backscatter mechanism
we developed to improve reliability
under the presence of external interference. The key idea is to use several commodity devices to
generate carrier signals at different frequencies that are then
backscattered simultaneously to multiple receivers.

\fakepar{Experiment setup}  We set up the experiment in our lab. We position
the backscatter tag \SI{1}{\meter} away from the carrier generators.
As carrier generators, we use three CC3200 WiFi radios~\cite{CC3200}, and program them 
to generate carrier signals at \SI{2412.3}{\giga\hertz}, 
\SI{2447.3}{\giga\hertz} and \SI{2472.3}{\giga\hertz}. We intentionally chose
these frequencies as they lie within the WiFi band of the interferer.
We position three receivers  \SI{8}{\meter} away from the carrier generators. To generate interference, we leverage another CC3200
radio to continuously generate WiFi traffic at maximum transmit power. We locate the interferer \SI{6}{\meter} 
away from both the  backscatter tag and the receivers.
As in the previous experiments, we calculate the BER.
%As a performance metric, similar to earlier experiment 
 
 \fakepar{Results}  We perform four runs of the experiment. In the first run we turn off the interferer.
 In the next three runs, we program the interferer to operate on WiFi channel 1, 7 and 13, respectively.
 Figure~\ref{unisoneval} shows the result.  In the first experiment without
 interference, we can receive transmissions from the tag on all
 the three frequencies, at a very low BER. For the next three runs, the figure shows a drop in BER of the 
 receiver whose frequency overlaps with that of the interferer.
 The figure shows that while there is 
 a significant  decrease in 
 BER at the receiver that operates on a frequency similar to the interferer, the other
 two receivers continue to receive at low BER. We precisely enable this diversity to improve reliability when 
 operating in interfered environments.

\subsection{ Avoiding interference }
\begin{figure}[!tb]
  \centering
  \includegraphics[width=\linewidth]{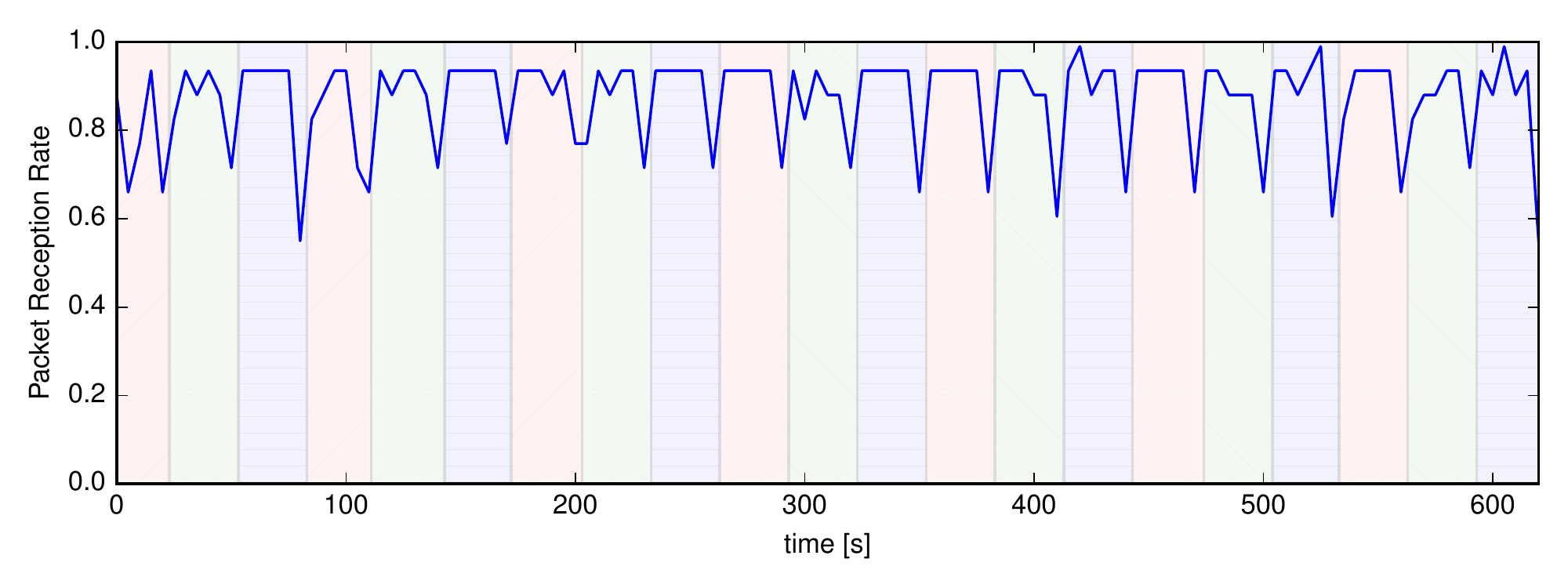}
  \caption{\capt{Changing carrier frequency to avoid interference}. Vertical color bands represent the
  operating frequency of the backscatter transmissions. A drop in PRR corresponds to periods of interference, causing the carrier generator to change frequency improving the PRR of the received backscatter transmissions.}
  \label{fig:avoidintf}

\end{figure}
\label{avoidintf}

Receivers commonly employed on backscatter tags are passive
envelope detectors which lack the necessary frequency
selectivity to perform carrier sensing operation~\cite{interscatter}.
Carrier sensing, however, is  important to
ensure that backscatter transmissions do not interfere with
ambient wireless traffic. To ameliorate the issue, we 
take advantage of the fact that carrier generators,
as well as receivers, usually are much more capable devices than the tags. 
Receiver and carrier generators can coordinate to first identify interference, 
and to change carrier frequency to ensure
weak backscatter transmissions avoid interference. We next demonstrate such a design:

\fakepar{Setup}  We program an SDR
to generate traffic imitating WiFi transmissions.
We program the SDR to change the frequency corresponding to
WiFi Channel 1, 7, 12 every \SI{30}{\second}. We keep the
backscatter tag and carrier generator about \SI{0.5}{\meter}
apart and program the receiver to respond to periods with high packet error rate
by sending instructions to the carrier generator
to change frequency. Note that this also induces a change in the frequency of the backscatter transmission itself, 
to which the receiver has to adapt. 
To avoid interference, the carrier changes frequency when notified by the receiver. In our experiments, the carrier selects a channel that will be interfered again when the interfering SDR changes frequency.

\fakepar{Result}
Figure~\ref{fig:avoidintf} demonstrates the result of the experiment. In
the figure, the  bands represent distinct transmission frequencies. 
We observe, as soon as there is a drop in the packet reception rate~(PRR) due to interference, the carrier generator
changes frequency (change in color), resulting in improvement in PRR, as the
backscatter transmissions are able to avoid the interfered channel.

\subsection{Comparison with CRFID}
\label{crfidcomp}
% vim: textwidth=72 spell spelllang=en_us
\label{crfid}
In this experiment we compare the performance of \system to CRFID tags
queried using a commercial RFID reader. We perform the experiment to
understand improvements in terms of  range.

\fakepar{Settings and metrics} We perform the experiment outdoors.
%in settings similar to Section~\ref{Outdoor}. XXX TV ref was broken
We use the Wireless
Identification and Sensing Platform (WISP)  as CRFID platform. WISP
has been widely used~\cite{wispcam,smith2006wirelessly} and
developed for close to a
decade~\cite{smith2006wirelessly}.  We use the present generation, and
the state-of-the-art WISP~5.0 for the experiments.  To query
the WISP tags, we use a commercial RFID reader (Impinj Speedway
R420~\cite{r420}, $\sim$ \$1600) equipped with a single \SI{9}{dBiC}
circular polarized antenna. We configure the reader to generate a carrier
signal of strength \SI{26}{dBm}, similar to the carrier strength used to
evaluate the \system reader.  We position the antenna and the WISP tags
approximately one meter above the ground.  As CRFID tags demonstrate an
asymmetry in the communication and energy harvesting
range~\cite{gummeson2010limits}, we externally power the WISP tags to
avoid being restricted by the energy harvesting range. 

In the same setting, to evaluate \system on \SI{2.4}{\giga\hertz},
we connect a \SI{9}{dBi} TP-LINK TL-ANT2409A antenna to the SDR. 
Due to the self-interference problem, we cannot use a monostatic
setup. We emulate the equivalent path loss of monostatic
configurations by keeping the carrier generator and the receiver
equidistant from the tag while maximising the distance between them.
We operate \system in low bitrate mode. 
We program both the WISP and \system to transmit with the minimum possible delay.
As a metric, we measure the achieved goodput.

\begin{figure}[!tb]
\centering
\includegraphics[width=\linewidth]{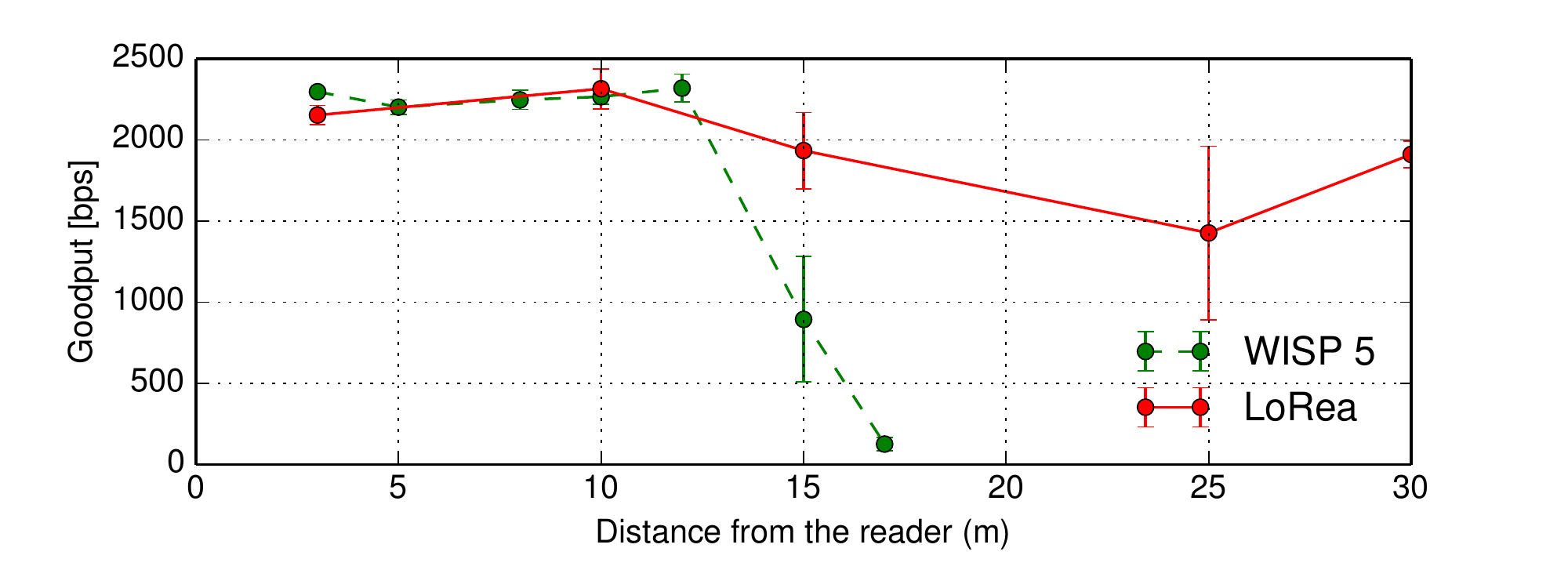}
\caption{\capt{Goodput comparison between WISP 5 and \system (outdoors).} WISP achieves
a maximum range of only 18 meters.}
\label{wispreadrate}
\end{figure}

\fakepar{Results} Figure~\ref{wispreadrate} shows that as the distance
between the WISP and the RFID reader increases, the achieved bitrate
drops significantly.  This is due to the SNR of the backscattered signal
decreasing at the reader and approaching the sensitivity level of the
reader.  We observe a maximum distance of approximately \SI{17}{\metre},
which is consistent with the 
maximum advertised range of the Impinj Speedway R420 RFID reader~\cite{r420}. 
Our architecture, in certain cases, achieves a range that
is one order of magnitude higher as compared to existing RFID readers.

The higher range achieved by our architecture  is  due to three reasons. First,
we  shift the weak backscattered signal away from the carrier
which reduces the interference, thereby improving the SNR. Second, we
use a radio which offers receiver sensitivity that is almost
\SI{20}{\deci\bel} higher
(approximately \SI{-104}{dBm})
compared to the \SI{-84}{dBm} the R420 reader offers, a typical sensitivity for
commercial RFID readers. Finally, most commercial RFID readers operate in a
monostatic configuration which, as we have discussed in
Section~\ref{subsec:carrier_generation}, limits the
achievable range significantly.

\fakepar{Interoperability } Our architecture, when operating at \SI{868}{\mega\hertz} 
is compatible and can be used together with the present generation of the WISP 5.0 CRFID tag with minor 
firmware modification to backscatter at an intermediate frequency.

\section{Proof-of-concept applications}
\label{applications}
\label{application_scenario}

%Existing radio technologies consume peak power consumption of tens of \SI{}{\milli\watt}s 
%to achieve communication range of hundreds of meters, which makes applications conceived using these
%radios reliant on batteries. On the other hand, state-of-the-art backscatter and CRFID systems 
%achieve low-communication range, and might also be expensive (existing CRFID readers cost
%upwards of USD 2000) which significantly increases the cost of deployment. 

%Our  architecture achieves communication range of hundreds of meters, while
%consuming \SI{}{\micro\watt}s of power at the backscatter tag, while also being low-cost~( lower than USD 70).
%As we had seen earlier, like all the recent backscatter systems the communication range is constraint on the 
%position of the tag from the receiver or the carrier generator. Inspite of this  constraint, our architecture can 
%improves vast number of applications conceived using
%CRFID platforms, and also enable novel applications
%difficult with existing state of the art. 

In this section, we present two proof-of-concept applications implemented
using \system which are challenging 
to realise with existing backscatter systems.

\subsection{Mobile Reader}
\label{mobilereader}

\label{parkinglotexp}
\begin{figure}
    \centering
    \includegraphics[width=\columnwidth]{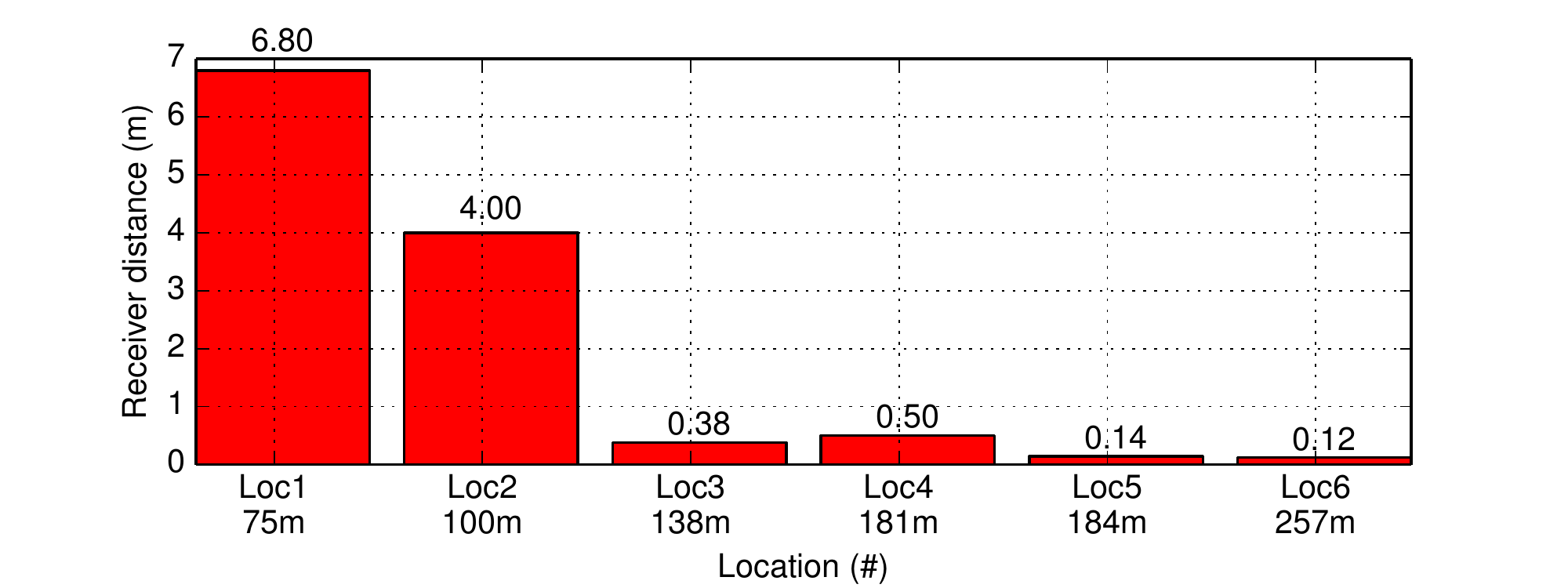}
    \caption{\capt{Receiving backscatter transmissions in parking space} (\SI{2.4}{\giga\hertz}). The farther the tag is from the carrier generator, the closer the reader has to be to the
    tag to receive.}
    \label{fig:parking}
\end{figure}
\label{mobilereader}

Mobile backscatter readers can be
useful for applications
in, for example,  libraries, offices, and
at manufacturing lines.
Existing backscatter readers, however,
usually combine carrier generation 
with reception, making them  bulky and power hungry which makes their operation difficult
in mobile scenarios.  

Our architecture can enable such applications, as
the bistatic mode delegates the more energy-expensive
tasks to the fixed infrastructure. This reduces the
power consumption of the receiver.
Decoupling the carrier generator, however,
introduces a new challenge: 
tags demonstrate varied
communication range, due to different
distances from the carrier generator.

To demonstrate this problem, we distribute backscatter tags
at six different locations 
in the parking space of the university. 
The backscatter tags are not in line-of-sight with the
carrier generator.
We find the maximum communication
distance between the tag and the receiver.
Figure~\ref{fig:parking} show that the range is
longer for tags closer to the carrier generator, while
for tags farther away the reader has to be close
to the tag to receive transmissions. 

In a concrete application scenario, one could deploy
several carrier generators as shown in Section~\ref{tonegeneration}.
Another option is to devise trajectories that allow the mobile reader 
to query the tags near the carrier generator from large distances and
tags farther away from the carrier generator from short distances.  
While we note that our architecture enables 
such applications due to its low power consumption, we leave
these issues to future work.

%\label{parkinglotexp}

\subsection{Sensors Embedded in the Infrastructure}
\label{concrete}
Embedding sensors in the infrastructure itself is an important challenge especially for
applications like structural health monitoring. These sensors measure parameters
like vibration, strain etc. and help improve the lifetime
of the infrastructure.  Making these sensors battery-free is important,
as they could be embedded within the structure and left unattended for long periods of time.
Existing attempts to embed CRFID sensors have resulted in very poor communication ranges (only a few meters), which severely
restricts their usage in real environments~\cite{alhaiderimoo}. Such a poor range is primarily due
to the large attenuation of RF signals while going through walls, coupled with the poor sensitivity levels of RFID readers.
The higher sensitivity of our receivers could enable 
\system to achieve high communication range. We explore this possibility next.

We place a tag in the basement of our building behind a thick concrete wall. Next, we place the carrier generator (\SI{868}{\mega\hertz}) with 
transmit power of \SI{24}{\decibel}m outside such that the wall separates the two. 
This scenario represents the worst
case scenario when compared to sensors embedded in the wall, as the backscattered
signal gets attenuated twice. 
We  find the distance up to which the receiver is able to
receive transmissions, as a function of the carrier generator's
distance from the tag. 

\begin{figure}
    \centering
    \includegraphics[width=\columnwidth]{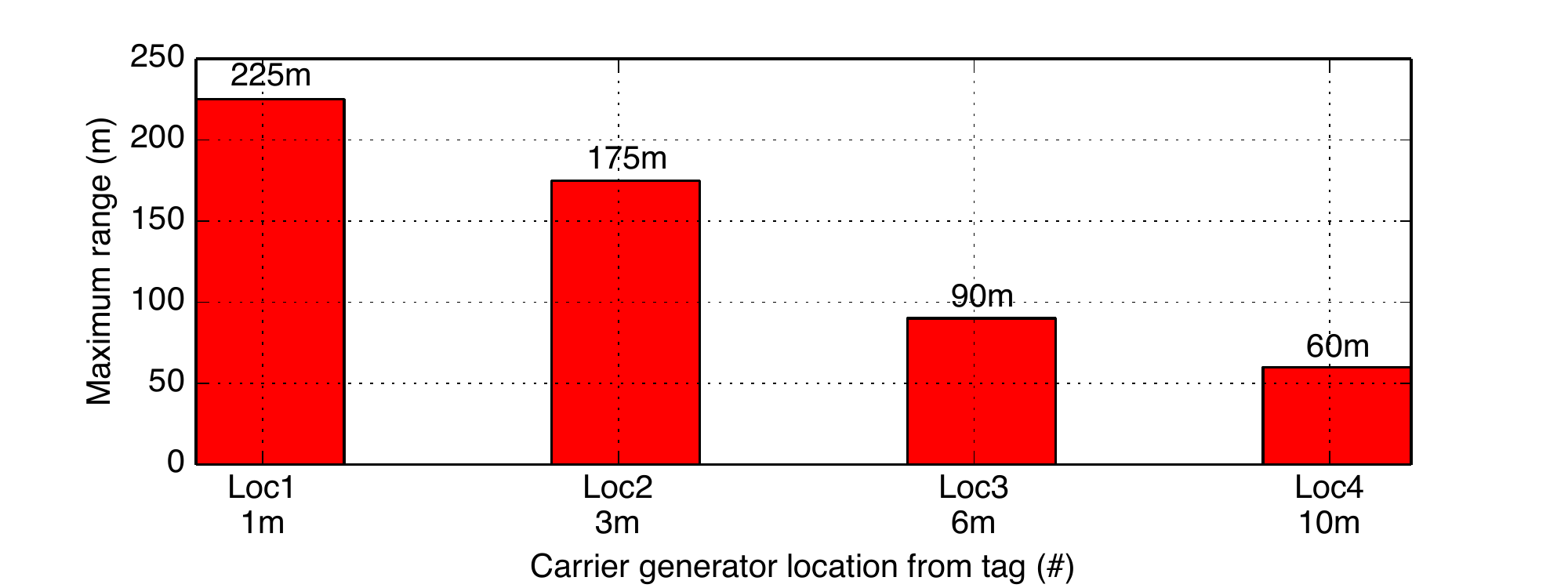}
    \caption{\capt{Embedding sensor in infrastructure} (\SI{868}{\mega\hertz}). Even in the presence of a thick concrete wall LoRea can receive backscatter transmissions hundreds of meters away from the tag.}
    \label{fig:concrete}
\end{figure}

The result of the experiment is shown in
Figure~\ref{fig:concrete}. The figure shows that at a distance of 
\SI{1}{\meter} between the tag and the carrier generator, we can achieve a communication range of 
\SI{225}{\meter}. Even when the carrier generator is \SI{10}{\meter} away,
we still achieve a significant communication range. We believe that our architecture 
takes a step to make these very important applications a reality.

\section{Discussion}
\label{discussion}
\fakepar{Commodity wireless devices as carrier generators} A key feature 
of \system's \SI{2.4}{\giga\hertz} architecture
is its ability to use existing wireless devices, such as WiFi routers and ZigBee hubs, to generate the carrier signal. 
%To achieve this capability, 
On these devices, \system uses the
%\system uses 
\emph{continuous carrier mode}
present to facilitate
regulatory compliance testing to generate the carrier signal. We performed a
brief survey and found access to this
mode in  many commercially-available WiFi~\cite{CC3200, esp8266},
ZigBee~\cite{CC2420,mc1319} and BLE radios~\cite{CYW20735,nrf51822}. 
Devices that use these transceivers  can generate a 
carrier signal  with only a minor modification to their firmware. For example, a vast number
of WiFi routers support the open source OpenWRT firmware~\cite{openwrt}.  
%without requiring hardware modification. 
OpenWRT enables
driver-level access to the WiFi transceivers
facilitating the configuration
required
to support the carrier
generation.

\fakepar{Supporting simultaneous transmissions from tags}  A crucial
requirement for backscatter readers is to support  
simultaneous reception from multiple backscatter
tags. This is particularly challenging
in our architecture due to the low data 
rate, which increases the probability 
%that
%backscatter tags send at the same time.
of collisions among backscatter transmissions.
Conventional backscatter tags 
transmit at the same frequency which results in frequent collisions
requiring mechanisms at the reader to separate the collided signals
and recover information~\cite{hu2015laissez,ou2015come}. 
However, such designs
increase the complexity and the cost.

In our architecture, we use heterodyning at the
backscatter tags to keep the carrier signal and backscatter
transmissions apart reducing self-interference.
Our recent work~\cite{ambujhotwireless} demonstrates that 
heterodyning also enables backscatter tags to operate
on distinct channels, thus enabling simultaneous  transmissions
without collisions.  We can build upon this to support
simultaneous transmissions without increasing the cost
and complexity of the reader required  by existing designs.
Due to the limited number of available
channels in the license free bands, a key challenge is to support a large number of
backscatter tags . We will explore this in the future.

\fakepar{Improving reliability of backscatter links}  Our architecture demonstrates
reliability comparable to state-of-the-art
backscatter systems~\cite{HitchHike,fmbackscatter,turbochargebackscatter,ambientbackscatter}.
However, the BER is higher than what is usually observed in 
conventional wireless systems especially at larger distances between the
tag and the carrier generator. We can improve 
reliability by building on our recent
work~\cite{ambujhotwireless} which uses 
bit spreading and forward error correction mechanisms to improve
the reliability of FSK backscatter transmissions at low SNRs.

\fakepar{Coordinating carrier generation} The continuous generation of carrier signals can pose a problem 
for the coexistence with other wireless devices and may not constitute an efficient use of the spectrum. 
A possible solution to this problem is to  coordinate the carrier generation so that 
carrier signals are only generated when tags should backscatter data. For example, the carrier
generator can be synchronised to generate the carrier signal with the wake-up period of the 
backscatter tags. %carriers generators to only
%it occurs only during the backscatter %transmissions, when it is actually needed. 
%The design of a protocol that takes care of the coordination is, however, outside of the scope of this paper.
The design of such a coordination protocol is, however, outside the scope of this paper.

\section{Conclusions}
\label{concs}

In this paper we have presented \system. \system departs from previous
CRFID designs in that it avoids the need for complex and expensive
self-interference cancellation. By decoupling carrier generation and reception,
\system also allows to leverage existing infrastructure for
generating the carrier and the use of highly sensitive narrow-band
receivers. \system is complemented by the novel design of a
backscatter tag that shifts and frequency modulates the carrier
signal while consuming \emph{$\mu W$s} of power.
\system achieves a  range beyond \SI{3.4}{\kilo\meter}  
when operating in the \SI{868}{\mega\hertz} band, and \SI{225}{\meter}
when operating in the \SI{2.4}{\giga\hertz} band
which is a significant improvement over the state of the art
in backscatter communication. The bistatic design of our architecture 
allowed to move complexity from the backscatter tag to the carrier generator 
and/or receiver, enabling several interesting applications as demonstrated in this paper.

%\system separates the backscatter transmissions 
%from the carrier by shifting the weak backscattered signal away from
%the carrier signal. Furthermore, \system decouples carrier
%generation from data reception which enables other devices such
%as sensor nodes and smartphones to generate the carrier,
%which improves range and scalability. In \system, receivers are cheap
%off-the-shelf radios with high sensitivity which leads to costs
%that are about one magnitude below that of commercial RFID
%readers. Putting these techniques together makes \system
%a low-cost but very efficient backscatter reader that achieves
%a range beyond \SI{200}{m} in line of sight, outperforming 
%state-of-the-art solutions.

%\section*{Acknowledgements}

\begin{acks} 

We  thank the anonymous reviewers and our shepherd Karthik Dantu for their insightful comments. 
This work has been funded by the Swedish Energy Agency~(Energimyndigheten).
\end{acks}

\bibliographystyle{acmref}

\bibliography{paper} 

\end{document}